\documentclass[journal]{IEEEtran}
\normalsize

\ifCLASSINFOpdf
  
\else
  
\fi

\usepackage{xcolor,soul,framed} 

\colorlet{shadecolor}{yellow}
\usepackage[pdftex]{graphicx}
\graphicspath{{../pdf/}{../jpeg/}}
\DeclareGraphicsExtensions{.pdf,.jpeg,.png}

\usepackage[cmex10]{amsmath}
\usepackage{array}
\usepackage{commath}
\usepackage{mdwmath}
\usepackage{mdwtab}
\usepackage{eqparbox}
\usepackage{url}
\usepackage{amsmath}
\usepackage{bbm}
\usepackage{yfonts}
\usepackage{dsfont}
\usepackage{ amssymb }
\usepackage[T1]{fontenc}
\usepackage[utf8]{inputenc}
\usepackage{authblk}
\usepackage[export]{adjustbox}
\usepackage{graphicx}
\usepackage{wrapfig}
\usepackage{lscape}
\usepackage{rotating}
\usepackage{epstopdf}
\usepackage{lipsum}
\usepackage{color,soul}
\usepackage{breqn}

\usepackage[numbers,sort&compress]{natbib}
\usepackage{amsthm,amssymb}

\usepackage{tabstackengine}
\setstackEOL{\cr}
\usepackage[utf8]{inputenc}
\newcommand{\argminF}{\mathop{\mathrm{argmin}}\limits} 

\newcommand{\argmaxF}{\mathop{\mathrm{argmax}}\limits} 

\begin{document}

\setlength{\textfloatsep}{1\baselineskip plus 0.2\baselineskip minus 0.5\baselineskip}

\title{Diffusive MIMO Molecular Communications:
Channel Estimation, Equalization, 
\\and Detection}

\author{S. M. Reza~Rouzegar and 
        Umberto~Spagnolini, \textit{Senior Member, IEEE}
\thanks{S. M. Rouzegar and U. Spagnolini are with Dipartimento di Elettronica,
Informazione e Bioingegneria, Politecnico di Milano, Milan, I-20133 Italy (e-mail: seyedmohammadreza.rouzegar@mail.polimi.it,  umberto.spagnolini@polimi.it).}
}

\maketitle

\newcommand{\vect}[1]{\boldsymbol{#1}}

\begin{abstract}
In diffusion-based communication, as for molecular systems, the achievable data rate depends on the  stochastic nature of diffusion which exhibits a severe inter-symbol-interference (ISI). Multiple-Input Multiple-Output (MIMO) multiplexing improves the data rate at the expense of  an  inter-link interference (ILI).  This paper investigates training-based channel estimation schemes for diffusive MIMO (D-MIMO) systems and corresponding equalization methods. Maximum likelihood and least-squares estimators of mean channel are derived, and the training sequence is designed to minimize the mean square error (MSE). Numerical validations in terms of MSE are compared with Cramér-Rao  bound derived herein. Equalization is based on decision feedback equalizer (DFE) structure as this is effective in mitigating diffusive ISI/ILI. Zero-forcing, minimum MSE and least-squares criteria have been paired to DFE, and their performances are evaluated in terms of bit error probability. D-MIMO time interleaving is exploited as an additional countermeasure to mitigate the ILI with remarkable performance improvements. The configuration of nano-transceivers is not static, but affected by a Brownian motion. A block-type communication is proposed for D-MIMO channel estimation and equalization, the corresponding time-varying D-MIMO MC system is evaluated numerically. 
\end{abstract}

\begin{IEEEkeywords}
Molecular communication, Diffusive Multiple-Input Multiple-Output (D-MIMO), Channel impulse response, Decision feedback equalizer, Maximum likelihood estimation, Zero forcing equalizer, Least squares error. 
\end{IEEEkeywords}

\IEEEpeerreviewmaketitle

\section{Introduction}

\IEEEPARstart{M}{olecular}  communication (MC) is a bio-inspired method based on transport of particles at nano-scale \cite{nakano2005molecular,farsad2016comprehensive,akyildiz2008nanonetworks,nakano2012molecular,akyildiz2011nanonetworks}. Conventional wave-field excitation based communication systems are impaired  by the wavelength and propagation at nano-scale. Using electromagnetic wave for nanomachines can be detrimental in some environments, such as inside a body where electromagnetic radiation can be harmful for health. Hence, MC is one preferred solution for communication among nanomachines to build nanonetworks that perform complex tasks \cite{akyildiz2015internet,nakano2014molecular,atakan2012body}. 

There are numerous potential applications envisioned for molecular communication such as medical application and communication between nanorobots. The continuous advances in nanotechnology, e.g. nanomachines and nanorobots, let us envision devices at nano-scale that are capable of computing and communicating \cite{cavalcanti2006nanorobot}. One of the applications of MC in medicine is artificial immune system \cite{okaie2014cooperative}, where many minuscule devices are injected to the body. Each tiny device is engineered for a specific task with limited functionalities. However, they can build a big nano-network to carry out  complex tasks such as targeted drug delivery \cite{nakano2012swarming} and cancer treatment \cite{couvreur2006nanotechnology}.

In MC, bio-nanomachines communicate through exchanging molecules through a liquid or gaseous environment. In fact, the simplest MC system needs a transmitter to send the information molecules, and a receiver to collect and process them. In the following, we briefly introduce each part of the diffusive MIMO (D-MIMO) MC system. 

\subsection{D-MIMO MC transmitter and receiver}
The transmitter is a bio-nanomachine which can be generated by genetically modified cells \cite{chen2005artificial}, artificial cells \cite{sasaki2010nanosensory} or also nanorobots. D-MIMO MC transmitter needs at least  a unit for  storing the information molecules and a unit for controlling the gates according to some input data. By controlling the gates opening time and size, one can encode the input signal to the different properties of molecules, such as their concentration \cite{kuran2011modulation}, number \cite{noel2014optimal}, type \cite{cobo2010bacteria}, and time of release \cite{farsad2016capacity,murin2017time}. Molecules can be any type of molecules according to the application and could be synthesized for drug delivery applications. Similarly, receivers can be bio-nanomachines or nanorobots. They should have at least a detection unit to sense the information molecules and a processing unit to decode the underneath data. 

\begin{figure}
  
  \includegraphics[width=2.5in,center]{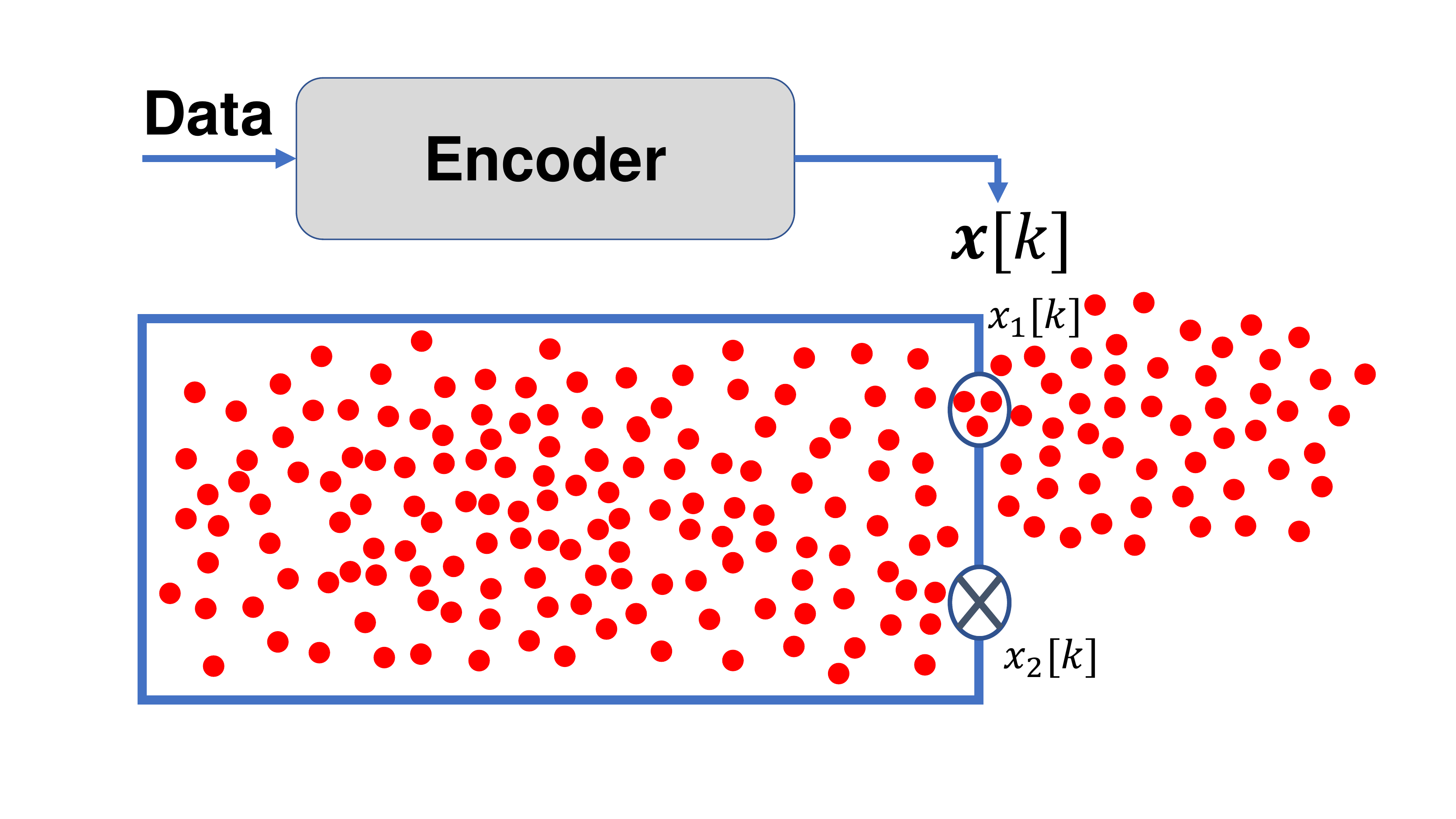}
  \caption{Transmitter of a $2\times 2$ D-MIMO system. The encoder unit controls opening time and aperture size of each gate, and therefore  the number of released molecules in each bit interval time. }\label{transmitter}
\end{figure}

In this paper, we have assumed that information is encoded in the number of molecules. Using ON-OFF key (OOK) signaling: the gate is open shortly for signaling bit one, and closed for signaling bit zero. Fig. \ref{transmitter} shows the simplified diagram of the MC transmitter for a $2 \times 2$ D-MIMO system. There are two independent gates where encoding unit controls their size and timing according to the input data. In other words,  the number of released molecules at each bit interval time is controlled at each gate according to the input data. 

\subsection{MC Channel}
Information molecules can be transported by different propagation mechanisms such as diffusion \cite{pierobon2010physical,arjmandi2013diffusion,jamali2016channel}, flow assisted diffusion \cite{noel2014optimal}, active transport using molecular motors and bacterial assisted propagation \cite{hiyama2010biomolecular,gregori2010new,enomoto2011design}. D-MIMO is restricted for simplicity to the setting where  information molecules diffuse toward the receiver using Brownian motion resulting from their collision with other molecules in fluid. 

 Diffusion can be modeled by Fick's laws of diffusion \cite{paul2014fick}. 
For an unbounded environment with a point source having an impulsive molecule release and transparent receivers at distance $d_{ij}$, the mean local concentration of molecules is \cite{mahfuz2014comprehensive,noel2014improving}: 
\begin{equation} \label{diff concentration}
\bar{\rho}_{ij} (d_{ij} ,t ) =
\frac{N}{(4\pi D \, t)^{3/2}}
\exp \big(-\frac{d_{ij}^2}{4D \, t} \,  \big) 
\end{equation}
where the mean local concentration of molecules is measured at the center of  $i$-th receiver at time $t$ after release of $N$ molecules by the $j$-th transmitter; $\mathrm{D}$ is the diffusion coefficient. The  mean local concentration vs time and space (\ref{diff concentration}) is for an ideal diffusion without any loss in Brownian motion of particles.

Diffusion is a stochastic phenomenon, the number of molecules varies in space and time, and the channel impulse response (CIR) $\bar{c}_{ij}(t)$ at distance $d_{ij}$ is defined as the \textit{expected} number of molecules at the receiver side at time $t$  after instantaneous release of $N$ molecules at $t=0$ \cite{jamali2016channel}
\begin{equation} \label{CIR integral}
\bar{c}_{ij}(t)=\iiint_{V^{Rx_i }}
\bar{\rho}_{ij} (d_{ij} ,t ) \,  \mathrm{d}x \, \mathrm{d}y \, \mathrm{d}z,
\end{equation}
where $V^{Rx_i }$ is the volume of the i-th receiver. In other words, $\bar{c}_{ij}(t)$ is the expected number of molecules of $j\rightarrow i$ link corresponding the release of $N$. The communication system is in a moving fluid and the environment is time-varying. In this setting, it is convenient to assume that CIR is constant in a certain channel coherence time $T_c$ (block-type communication) and thus within $T_c$ the CIR variations are negligible enough to impair the receiver performance. In D-MIMO, we assume the number of molecules at any receiver follows the \textit{Poisson distribution}  according to ref. \cite{jamali2016channel,arjmandi2013diffusion,mosayebi2014receivers}, with the mean number of molecules $\bar{c}_{ij}(t)$.

Diffusion model is a convenient abstraction used here for the discussion of diffusive MIMO systems. However, practical diffusive channels are affected by boundary, possibly absorbing (e.g., enzyme recombination makes the communication link lossy) or reflecting, with a non-homogeneous coefficient D, and other molecule loss. Further, transmitters are not point-like release of molecules, and receivers influence unavoidably the diffusion while collecting the information molecules. In this paper, all these application-specific complexities are not considered but still the stochastic nature of the propagation is considered as detailed in Section III.

\subsection{Challenges, Related Works and Contributions}
One of the main challenges of MC is to deal with the long tail of diffusive propagation that causes severe and peculiar inter-symbol-interference (ISI) which is stochastic for the effect of diffusive channel. To cope with stochastic ISI one can increase the bit interval time to mitigate the ISI effect, but the exploitation of the spectrum efficiency justifies the optimization of the bit interval time and incorporate few channel taps into CIR due to the ISI \cite{kim2014symbol}. Even if one optimizes the bit interval time, the  slow nature of diffusion makes the error-free datarate quite low. Using multiple-input multiple-output (MIMO) technique is a widely investigated solution in communication systems to address this limit and that can be adopted for MC \cite{koo2016molecular,meng2012mimo}. 

\begin{figure} 
  \includegraphics[width=3.5in]{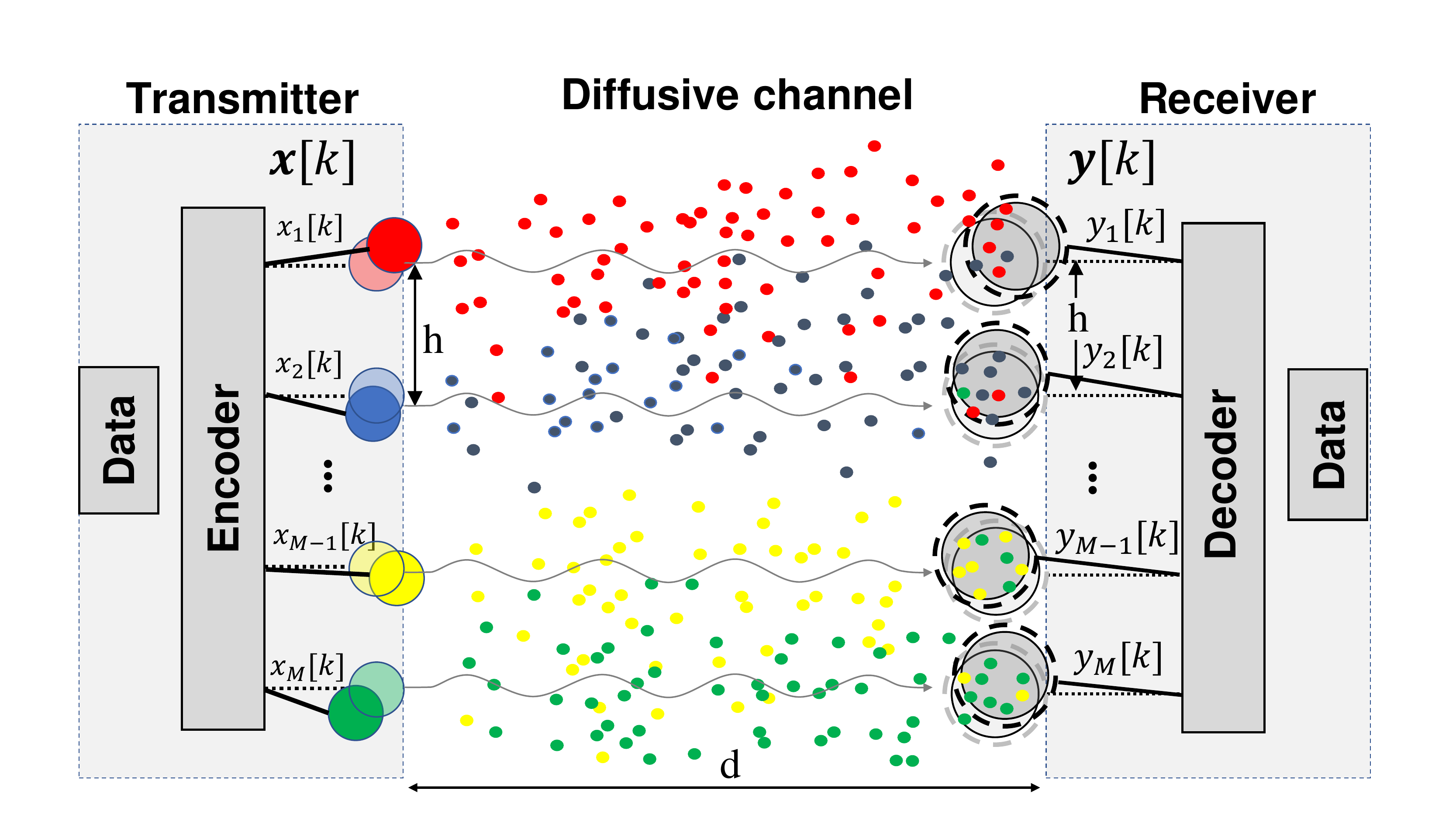}
  \centering
  \caption{ Topological model for $M\times M$ D-MIMO system. The $M$ transmitters $(Tx_1,...,Tx_M)$ release the same type of molecules (circle shape), and molecules here have different colors according to the corresponding transmitter to visualize the interference phenomena. Receivers $(Rx_1,...,Rx_M)$ counts the  molecules from the paired transmitter as well as non-paired transmitters called ILI. Random movements of the transmitter and receiver around their nominal position make the diffusive channel to be time-varying. }\label{N_N}
\end{figure}

The focus of this paper is to introduce a Diffusive MIMO (D-MIMO) MC system based on block-type communication in time-variant diffusive channels. In detail, a D-MIMO system model is proposed  (Section \ref{System model})  which accounts for random channel taps due to the diffusive nature which is not common in conventional communication systems. Movement of the communication system (e.g., in fluid) makes the diffusive channel time-varying, and the CIR is estimated at the beginning of each block by signaling a known training sequence over the D-MIMO channel. Maximum likelihood (ML) and least squares (LS) CIR estimators are proposed and their performances are compared with the Cramér-Rao bound (CRB). Herein, one-shot detectors are adopted for their low complexity. Decision feedback equalizer (DFE) is used to cancel the mean interference due to the long memory of the D-MIMO channel. Zero forcing (ZF) and minimum mean square error (MMSE) criteria are exploited in DFE to mitigate the D-MIMO inter-link interference.  Since performance of the D-MIMO system is limited by the inter-link interference (ILI), D-MIMO time inter-leaving (TIL) encoding is proposed. TIL encoder uses the time diversity between different gates of each MC transmitter to avoid interference, and thus reducing the error probability.

MC is widely investigated  in the literature but diffusive MIMO MC communication has  attracted some attentions only recently in \cite{koo2016molecular,meng2012mimo}. The overall contribution of this paper can be summarized below: 
\begin{itemize} 
\itemsep0em 
\item 
This paper proposes a training-based channel estimation  based on a matrix model for a  $M\times M$ MC systems shown in Fig. \ref{N_N}. The steps of V. Jamali et. al. \cite{jamali2016channel} have been extended to D-MIMO channel estimation (Section \ref{D-MIMO Channel Estimation}) by accounting for the inter-link diffusive-type interference. In a previous paper \cite{rouzegar2017channel}, single-line D-MIMO channel estimation was adopted, while here we generalize the approach to multi-line D-MIMO channel estimation.  
\item 
A method for designing training sequences is presented in this paper by minimizing the  Cramér-Rao bound (CRB). It differs from the method used in \cite{jamali2016channel} for the search strategy to cope with M transmitting gates, to reduce  the complexity that otherwise would be exponential.    
\item 
Severe interference in D-MIMO MC justifies the multi-channel decision feedback equalizer (DFE) to reduce the mean interference. DFE has been considered in \cite{kilinc2013receiver}, \cite{tepekule2015isi} and \cite{ahmadzadeh2014analysis} for SISO and Multi-Hop diffusive MC, respectively. DFE is extended here to D-MIMO MC system to reduce the mean interference, leaving the stochastic component to augment the noise.
\item 
Equalization and detection schemes for MC single-transmitter single-receiver are in \cite{noel2014optimal,mosayebi2014receivers,jamali2016non,kilinc2013receiver,meng2014receiver,li2016local,tepekule2015novel}. Authors of \cite{noel2014optimal} derived the maximum
likelihood sequence detector using Viterbi algorithm for known CSI.   MAP and MSSE sequence detectors are in \cite{kilinc2013receiver}  with a complexity that grows with sequence length. One-shot detectors are preferred here for their low complexity in nano-networks. In \cite{meng2014receiver} authors showed that the performance of one-shot detectors can reach those of the maximum likelihood sequence detector, but with perfect a-priori information and only for large bit time intervals (low interference environment). Assuming Gaussian approximation for Poisson diffusion \cite{marcone2017gaussian}, authors of \cite{koo2016molecular} investigated  linear zero-forcing  (ZF) criteria for a MC MIMO setting. Herein, ZF and MMSE criteria are employed in non-linear DFE equalizers for D-MIMO. Moreover, least squares DFE (LS-DFE) is introduced for decoding the data affected by Poisson stochastic term with exponential computational complexity in $M$.  
\item 
Performance of the D-MIMO system is dominated by the inter-link interference (ILI), thus D-MIMO time diversity is exploited at the transmitter to reduce the error probability. Particularly, D-MIMO time inter-leaving (TIL) encoder is introduced at the transmitter to modulate molecules release. Authors in \cite{noel2014overcoming} evaluated the effect of asynchronous multi-user interference for SISO MC system and showed that ILI decreases when there is an offset time between different transmitters. Here, D-MIMO TIL encoder controls the offset-time between the gates of a transmitter to reduce the error probability. By employing D-MIMO time and space diversity, the goal is to maximize the datarate over the allocated spatial arrangements of transceivers.
\item
In MC systems the diffusive channel is time-varying. In ref.  \cite{ahmadzadeh2017diffusive} authors  modelled the transceivers mobility by Brownian motion characterized by a relative diffusion constant for the whole system. Authors of \cite{ahmadzadeh2018stochastic} studied the statistical characterization of time-varying CIR for a single-link  MC channel.  Transceivers' movement has been modelled by three-dimensional random walk, and information is exchanged through collisions in \cite{guney2012mobile}.  Here, transceivers' mobility is modelled by particle-based Brownian motion and unlike ref.  \cite{guney2012mobile} the information is exchanged via diffusion of signaling molecules. Similarly to time-varying wireless communication systems, here we use a block-type protocol where the CIR is estimated at the beginning of each block.  D-MIMO context herein is time-varying, transceivers are randomly moving during channel estimation and detection phase. Numerical analysis of  time-varying D-MIMO  channel is done to show the impact of block-sizes on the overall MC system performance. 
\end{itemize}

The overall paper is organized as follows: Section \ref{System model} introduces the algebraic model for a $M\times M$ D-MIMO system. Section \ref{D-MIMO Channel Estimation} presents the CIR estimation for D-MIMO Poisson MC channel, maximum likelihood (ML) and least squares (LS) CIR estimators are proposed and their performances are compared with the Cramér-Rao bound (CRB). Section \ref{receiver Design} proposes equalization and  detection techniques for the D-MIMO system. In Section \ref{Interference Mitigation Technique}, D-MIMO time interleaving encoding technique is investigated to reduce the ILI effect and increase the transmission efficiency. Finally in Section \ref{Performance Evaluation and parameter study}, the numerical analysis of the block-type MC communication is provided and the impact of parameters on performance of the D-MIMO system is evaluated.

\section{System model } \label{System model}
 
We consider a $M\times M$ D-MIMO system for MC shown in Fig. \ref{N_N}. The system consists of $M$ pairs of transmitters labeled as $Tx_j $ and receivers $Rx_i$, where $i,j \in \{ 1,2,3...M\} $.
We assume that  all transmitters emit the same type of molecules. Each transmitter emits a known number of molecules $N$ at the beginning of each bit interval time $T_{int}$ to signal bit 1, and stays silent to signal bit 0. The $N$  molecules diffuse in the environment and some of them reach the set of $M$ receivers.
Transmitters and receivers are not fixed in their position but are moving randomly in the fluid where the signaling molecules diffuse, so CIR changes over the time. We assume a block-type communication and we estimate the CIR at the beginning of every block by sending a properly designed training sequence (Sec. \ref{D-MIMO Channel Estimation}). The estimated CIR is then used for equalization and detection  over the rest of the block (Section IV) while D-MIMO channel response continuously varies thus outdating along the block.

Transmitters modulate the molecules density using concentration shift keying (CSK) and the receivers count the number of molecules at the time of sampling. As customary, sampling time is set to the peak concentration time ($\tau_{max}$) where numbers of molecules received  from the corresponding transmitters are at their maximum. The counted number of received molecules is a random variable for diffusion mechanism and it is further impaired by the inter-symbol-interference (ISI) molecules of the corresponding transmitter. Similarly, inter-link interference (ILI) is due to the molecules from the current and previous samples of the non-corresponding transmitters gathered on the time reference of the receiver (delayed for the causality). Specifically, we can eliminate the ISI and ILI by making the bit interval time large enough and putting each pair of transceivers far enough from the other pairs. However, this situation may not be feasible in MC systems and it leads to a loss of datarate per unit of space. Hence, MC system should be designed to mitigate ISI and ILI  by proper equalization and detection. The observed number of molecules at time interval $k$ and $i$-th receiver for channels' length $\mathrm{L}$ is
\begin {equation}\label{first equation}
y_i[k]=\sum_{j=1}^{M}\sum_{\ell=0}^{L-1} c_{ij}[\ell,k]
x_j[k-\ell]  +v_i[k]
\end{equation}
where $c_{ij}[\ell,k]$ is a random variable accounting for the number of molecules observed at time $k$ by the receiver $Rx_{i}$ from transmitter $Tx_{j}$ due to the release of $N$ molecules at  time interval $[k-\ell]$. Case $i=j$ refers to the paired transmitter-receiver, otherwise it refers to the ILI. $x_j[k] \in \{0,1\}$ is the transmitted symbol at  time interval $k$ from $Tx_j$ activating, or not, the emission of the $N$ molecules. The number of molecules  $c_{ij}[\ell ,k]$ is modeled as a Poisson random variable with  mean value $\bar{c}_{ij}[\ell]$: $c_{ij} [\ell,k] \sim Poiss \, (\bar{c}_{ij} [\ell]) $. Additionally,
$v_i[k]$ is the number of noise molecules detected at the receiver $i$ at time interval $k$. These noise molecules might originate from the  channel taps of all transmitters that are not accounted by the L-taps model  (\ref{first equation}) and any external source. Thus, it is approximated as Poisson variable with mean: $\bar{v}_i$: $v_{i}[k] \sim Poiss \, (\bar{v}_{i})$ \cite{arjmandi2013diffusion,jamali2016channel,pierobon2011diffusion}.  
Even if the diffusive channel relationship (\ref{first equation}) is nonlinear because output $y_i[k]$ is not linearly dependent on the input, but via Poisson distribution, its mean value with respect to the  channel diffusivity is linearly dependent on the bit symbols $x_j[k]$  
\begin {equation}\label{mean1 equation}
\bar{y}_i[k]= \mathbb{E}_{c|x} \, \, \{ y_i[k] \}=
\sum_{j=1}^{M}\sum_{\ell=0}^{L-1} \bar{c}_{ij}[\ell] 
x_j[k-\ell]  + \bar{v}_{i},
\end{equation} 
to avoid edge effect due to the ISI, we employ $y_i[k]$ for $ k\geq L$. Analysis of the mean value of Poisson distributed (\ref{mean1 equation}) for D-MIMO is described below.

To simplify the reasoning, let the channel be memoryless ($L=1$), the $M\times M$ D-MIMO system is 
\begin{equation}\label{ISI-free relation} 
\bar{\vect{y}}[k]=\vect{\bar{C}}[0] \, \vect{x}[k] + \bar{\vect{v}},
\end{equation}
where $\vect{x}[k]=[x_1[k],x_2[k],..., x_M[k]]^T$ is  binary data at time interval $k$, $\bar{\vect{y}}[k]= [\bar{y}_1[k],\bar{y}_2[k],..., \bar{y}_M[k]]^T$ is the expected number of molecules at $M$ receivers at $k$-th time interval and $\bar{\vect{v}}=[\bar{v}_1, \bar{v}_2, \dots, \bar{v}_M ]^T$ is the corresponding noise molecules. Defining  $\vect{\bar{C}}[\ell]$ as the $\ell$-th tap of mean CIR of a D-MIMO channel
\begin{equation} 
\setstacktabbedgap{2pt}
\vect{\bar{C}}[\ell]=  \parenMatrixstack{
\bar{c}_{11}[\ell] & \bar{c}_{12}[\ell] & \dots & \bar{c}_{1M}[\ell]  \cr
\bar{c}_{21}[\ell] & \bar{c}_{22}[\ell] & \dots & \bar{c}_{2M}[\ell]  \cr
\vdots & \vdots & \ddots &  \vdots \cr
\bar{c}_{M1}[\ell] & \bar{c}_{M2}[\ell] & \dots & \bar{c}_{MM}[\ell] 
},
\end{equation} \label{C0 matrix}
the matrix $\vect{\bar{C}}[0]$ denotes the special case of memory-less channel.

Extending now to diffusive channel with $L$ memory taps, the D-MIMO relation (\ref{mean1 equation}) is 
\begin{equation}\label{D-MIMO k relation}
\bar{\vect{y}}[k]=\vect{\bar{C}} \, \vect{X}[k] 
\end{equation}
where $\vect{\bar{C}}=[\vect{\bar{C}}[0],\vect{\bar{C}}[1], \dots , \vect{\bar{C}}[L-1],\bar{\vect{v}}]$ is a $M\times (ML+1)$ matrix denoting the global D-MIMO channel matrix, including the channel memory taps, and $\vect{X}[k]=[\vect{x}^T[k],\vect{x}^T[k-1],...., \vect{x}^T[k-L+1],1]^T$ is the corresponding  $(ML+1)\times 1$ vector of the transmitted sequence at time $k$ completed by the $L-1$ previous time intervals. Notice that noise vector $\bar{\vect{v}}$ augments the global channel matrix $\vect{\bar{C}}$ as this simplifies the analytical settings. Model (\ref{D-MIMO k relation}) is for the expected number of molecules at the $k$-th time interval of the $M$ receivers. The global D-MIMO relation at all $K$ time intervals can be expressed as (dimensions are in subscript):
\begin{equation}\label{Global MIMO relation}
\underset{M\times (K-L+1) }{\vect{\bar{Y}}}=\underset{M\times (ML+1) }{\vect{\bar{C}}} \,\,\, \underset{(ML+1)\times (K-L+1)}{\vect{X}}
\end{equation}
where $\vect{\bar{Y}}=[\bar{\vect{y}}[L],\bar{\vect{y}}[L+1],\dots ,\bar{\vect{y}}[K]]$ denotes the global receiver matrix, and $\vect{X}=[\vect{X}[L], \vect{X}[L+1], \dots , \vect{X}[K] ]$ is the global convolutional matrix on $\mathrm{L}$ memory taps.

According to the above notations, individual output (\ref{mean1 equation}) can be rewritten compactly as:
\begin{equation}\label{y j receiver}
\bar{y}_i[k]=\vect{\bar{C}}_i \, \vect{X}[k]
\end{equation}
where $\vect{\bar{C}}_i= \vect{\bar{C}} (i,:)$ is the $i$-th row of the global channel matrix $\vect{\bar{C}}$ and it collects all mean channel taps which are received at $i$-th location, $\vect{X}[k]$ is the $(k-L+1)$-th column of the global matrix $\vect{X}$. The observed number of molecules by the $M$ receivers is a set of Poisson random variables with mean $\vect{\bar{Y}}=\vect{\bar{C}} \vect{X}$:
\begin{equation}\label{Poiss Y}
\vect{Y}=\vect{C} \vect{X},
\end{equation}
where $\vect{C}=\mathrm{Poiss} \,(\vect{\bar{C}})$ and each entry of $\vect{Y}$ is Poisson distributed with mean equal to the corresponding entry of  $\vect{\bar{Y}}$.

\section{D-MIMO Channel Estimation}\label{D-MIMO Channel Estimation}
Practical diffusion is far more complex than releasing $N$ molecules over an ideal diffusive channel. In real environments, system parameters are not known a-priori and transmitter is not able to instantaneously release exactly $N$ molecules. Moreover, diffusive channel varies over time, other physical and chemical phenomena could enter into play such as some pre-existing enzymes in the environment, molecules degradation and recombination, just to mention few \cite{noel2014improving,ahmadzadeh2016comprehensive}. In these settings CIR  should be estimated continuously over time. This justifies the use of block-type communication where training sequences are transmitted at the beginning of each block, and receivers estimate the CIR by knowing the training sequence from the observed number of molecules. Block-length $B$ is chosen such that within each block, the channel variation would be negligible.

In this section, we employ the D-MIMO system model and we use $\vect{S}$ instead of  $\vect{X}$ to  denote the global training sequence  matrix which is known at the receiver. Therefore, the D-MIMO relation (\ref{Global MIMO relation}) for channel estimation is
\begin{equation}
\bar{\vect{Y}} = \bar{\vect{C}} \, \vect{S}
\end{equation}
where $\vect{S}[k]=[\vect{s}[k],\vect{s}[k-1],\vect{s}[k-L+1],1]^T$. To avoid edge effect due to the ISI, we employ $y_i[k]$ for $ k\geq L$ in CIR estimation and the $K-L+1$ samples are used for CIR estimation of the $i$-th receiver.

Due to the independence, the probability density function (PDF) of all observations at all receivers $\vect{Y}$ are the product of the Poisson distribution of each observation at each receiver 
\begin{equation}\label{PDF}
f_{\vect{Y}} \, (\vect{Y}|\vect{\bar{C}},\vect{S}) = \prod_{k=L}^K \prod_{i=1}^{M}
\frac{(\vect{\bar{C}}_i \,\vect{S}[k])^ {y_{i}[k]} \, \exp (- \vect{\bar{C}}_i \,\vect{S}[k])}
{y_{j}[k] \,!}
\end{equation}
where $\vect{S}[k]$ is the $(k-L+1)$-th column of the training sequence matrix $\vect{S}$ and $y_i[k]=Y_{ik}$ is the $Y[i,(k-L+1)]$ entry of $\vect{Y}$. The log-likelihood function can be written as 
\begin{multline} \label{log global}
\mathcal{L} _{\vect{Y}} \, (\vect{Y}|\vect{\bar{C}},\vect{S}) = \sum_{i=1}^{M}\sum_{k=L}^{K} \big [   -\vect{\bar{C}}_i \,\vect{S}[k] + y_{i}[k] \, \ln \,(\vect{\bar{C}}_i \,\vect{S}[k] )\\ -\ln (y_{i}[k] \,!)\,\big ],
\end{multline}
that is instrumental for deriving Cramér-Rao Bound and maximum likelihood estimation as detailed below. Recall that estimating the CIR implies to estimate the mean noise term $\vect{\bar{v}}$.

\subsection{Cramér-Rao Bound}
The Cramér-Rao bound (CRB) sets the lower bound on the covariance of any unbiased estimator of a deterministic parameter. Let $\hat{\bar{\vect{C}}}$ be the unbiased estimator of $\bar{\vect{C}}$, the CRB sets the bound of the covariance
\begin{equation}\label{CRB inequality}
\mathrm{cov}  (\hat{\bar{\vect{C}}}) \succeq  \vect{I}^{-1} (\bar{\vect{C}}),
\end{equation}
where $ \vect{I} (\bar{\vect{C}})$ is the Fisher information matrix of  $\bar{\vect{C}}$ evaluated as Eq. (7) in ref. \cite{jamali2016channel}
\begin{equation} \label{Fisher}
\vect{I} (\bar{\vect{C}})= \mathbb{E} _{\vect{Y}} \{ 
- \frac{\partial^2 \,\,\mathcal{L} _{\vect{Y}} \, (\vect{Y}|\vect{\bar{C}},\vect{S})}{ \partial^2 \vect{\bar{C}}} \}, 
\end{equation}
Therefore, the CRB of the CIR is 
\begin{equation}\label{CRB}
\mathrm{CRB_{MIMO}}= \mathrm{tr}\{\vect{I}^{-1} (\bar{\vect{C}})\}=
tr\bigg\{ \bigg[\sum_{i=1}^{M} \sum_{k=L}^{K} \frac{\vect{S}[k]\vect{S}^T[k]} {\vect{\bar{C}}_i \, \vect{S}[k]  } \bigg] ^{-1}
\bigg\},
\end{equation}
and $\mathrm{CRB_{MIMO}}$ sets the reference bound for MSE of CIR estimators.
\subsection{Maximum Likelihood CIR estimator}
Maximum likelihood (ML) D-MIMO CIR estimator finds the positive values of $\bar{\vect{C}}$ which maximize the likelihood of observations $\vect{Y}$ 
\begin{equation}\label{ML}
\hat{\bar{\vect{C}}}^{\mathrm{ML} } = \argmaxF_{{\bar{\vect{C}} \geq 0}}  \,  f_{\vect{Y}} \, (\vect{Y}|\vect{\bar{C}},\vect{S}) =\argmaxF_{{\bar{\vect{C}} \geq 0}}  \,  \mathcal{L} _{\vect{y}} \, (\vect{Y}|\vect{\bar{C}},\vect{S}) 
\end{equation}
where the log-likelihood function is defined in Eq. (\ref{log global}). 

The ML estimate of the CIR for the D-MIMO channel is obtained by solving a set of non-linear equations similarly to Eq. (13) in ref. \cite{jamali2016channel}: 
\begin{equation} \label{ML equations}
\sum_{i=1}^{M} \sum_{k=L}^{K} \big [ \frac{y_{i}[k]\vect{S}[k]}{\vect{\bar{C}}_i\vect{S}[k]} -\vect{S}[k] \,
\big ] = \vect{0}
\end{equation}
for positive entries of $\vect{\bar{C}}_i$. In other words, we are searching for the channel entries $\vect{\bar{C}}_i$ where the expected ($\vect{\bar{C}}_i\vect{S}[k]$) and observed number of molecules ($y_{i}[k]$) coincides. Even if entries of $\vect{\bar{C}}$ are positive valued (indicated as $\vect{\bar{C}} \ge 0$), numerical methods could yield a negative value for some elements of $\vect{\bar{C}} $. Sub-optimal solution is to constraint to zero all the negative entries of the estimated CIR.  This heuristic approach was adopted for single link MC in \cite{jamali2016channel} with a negligible loss of performances compared to the optimal ML (\ref{ML}). Extension to D-MIMO of this method preserves the property of optimality as proved later in Section V,
this sub-optimal solution of (\ref{ML equations}) is highly preferred in D-MIMO channels due to its simplicity. 
\subsection{Least Squares CIR estimator}

The Least-Squares (LS) method chooses $\vect{\bar{C}}$ which minimizes the sum of the square errors compared to the observation vector $\vect{Y}$: 
\begin{equation}\label{arg LS}
\hat{\bar{\vect{C}}}^{\tiny LS} = \argminF_{{\bar{\vect{C}} \geq 0}} \, \big \| \vect{\vect{Y}- \vect{\bar{C}} \vect{S} } \big\| ^2.
\end{equation} 
The unconstrained LS estimate of the CIR for D-MIMO channel is
\begin{equation} \label{LS Estimator}
\hat{\bar{\vect{C}}}^{\tiny LS} = \big[  (\vect{S} \, \vect{S}^T) ^{-1}   \, \vect{S} \, \vect{Y}    \big],
\end{equation}
but the constrained optimization problem is with $\bar{\vect{C}}   \geq 0$ for entries. 
Sub-optimal solution for (\ref{arg LS}) is again by setting all the negative entries to zero \cite{jamali2016channel}. In spite of simplicity, for $K$ large there exist a stationary point, and for small $K$ the performance loss is very negligible  as shown numerically in Section \ref{Performance Evaluation and parameter study}. We refer to this sub-optimal solution for D-MIMO system as LS-CIR estimator. 
 
\subsection{Training Sequence Design}
Accuracy depends on the training sequence (\ref{CRB}), here we exploit this dependency to design training sequences for D-MIMO CIR estimation. From $\mathrm{CRB_{MIMO}}$ (\ref{CRB}) that depends on the global training sequence matrix $\vect{S}$, one can find one (or more) $\vect{S}$ such that $\mathrm{CRB_{MIMO}}$  is minimized. Let \{$\vect{s}_1, \vect{s}_2, \dots, \vect{s}_M $\} be a training sequence with length $K$ at each transmitter. For a $M\times M$ D-MIMO system, one has to design $M$ different training sequences to minimize the $\mathrm{CRB_{MIMO}}$. However, in practice we do not need to design $M$ distinct training sequences because ILI for far transmitters is negligible, and thus  we neglect their interference channels but consider them as an augmented noise source in $\bar{v}_i$. Exhaustive search should be done among all possible  combinations of binary valued training sequences and choose those that minimize the corresponding $\mathrm{CRB_{MIMO}}$ (\ref{CRB}). Complexity in searching among all possible combinations is  $\mathcal{O} (2^{M\times K})$ and so even for small length $K$ it is extremely time-consuming. Therefore, we  employ a search algorithm to discard unfavored combinations and thus reduce the number of searches dramatically \cite{rouzegar2017channel}.

First, one chooses a short and suitable training sequence for each transmitter with smaller length $K_1$, and then the $M$ sequences each are concatenated to build the training sequences with length $K$.  Performance degradation due to the concatenating step is proved to be negligible for $M=1$ \cite{jamali2016channel}, and similarly for $M>1 $ as numerically validated  in Section \ref{Performance Evaluation and parameter study}. Second, the training sequences should be molecularly efficient by minimizing the fraction of molecules used for channel estimation (recall that molecules are released only when $s_j[k]=1$), and it should be constrained so that each gates should not be silent for many consecutive intervals. In detail, for a training sequence of length $K$, we consider sequences with maximum $K/2$ ones  and maximum $(L+1)$ consecutive zeros. Consequently, the D-MIMO system transmits at most $MNK/2$ molecules for each channel estimation. Let $\mathcal{S}$ be the sets of all possible training sequences that meet the above criteria, a semi-exhaustive search is employed to find the optimized training sequence:
\begin{equation}\label{sequence}
[\vect{s}_1,\vect{s}_2,...,\vect{s}_M]=
\argminF_{\vect{s}_i\in \mathcal{S}} \{ \mathrm{CRB_{MIMO}} \}.
\end{equation}
Accuracy of CIR estimation depends on the training sequence length $K$ and there is a trade-off between accuracy and molecular efficiency. In Section \ref{Performance Evaluation and parameter study}, all the training sequences are designed according to the method above.

 \section{D-MIMO Equalization and Detection} \label{receiver Design}
 
\begin{figure}
  \includegraphics[width=3.2in]{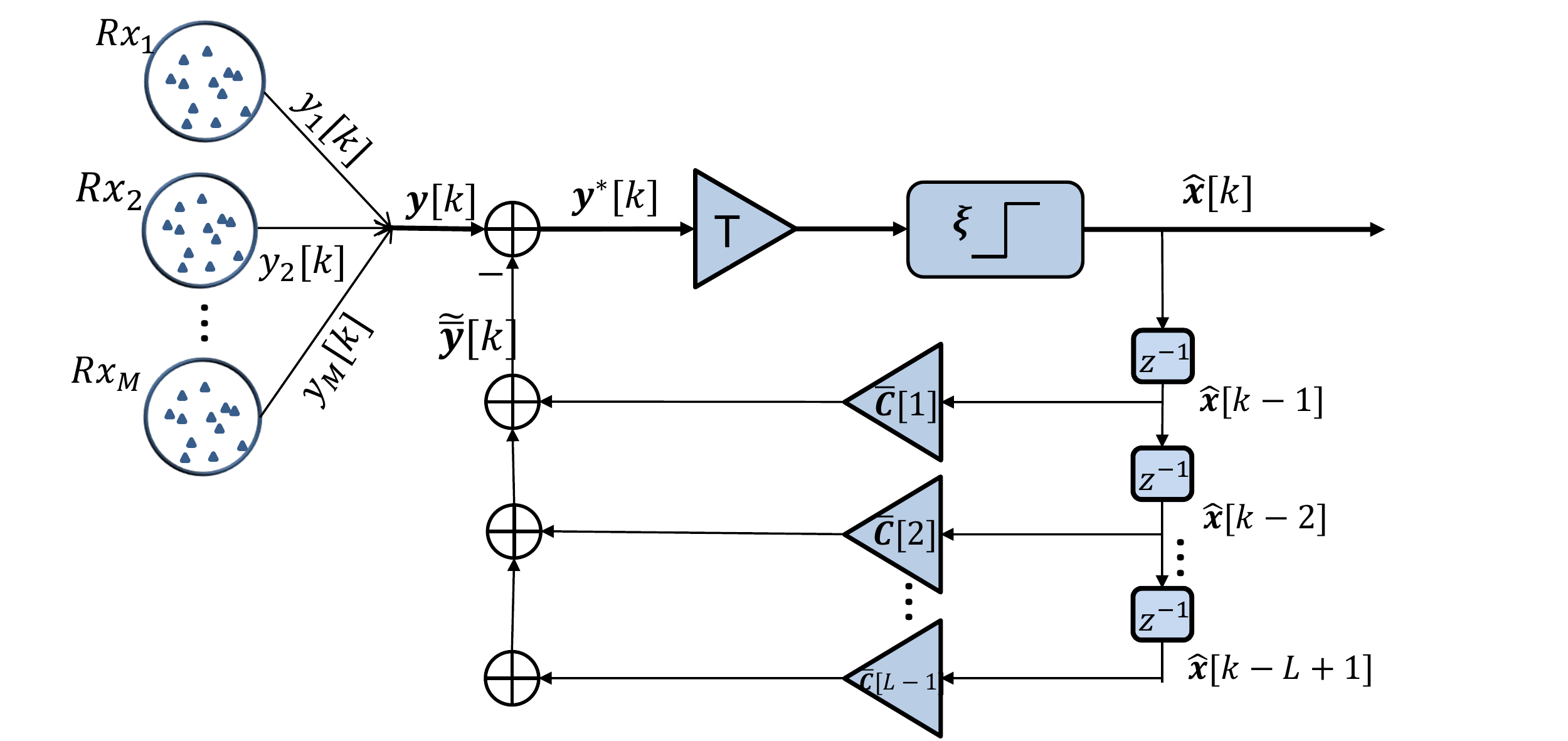}
  \centering
  \caption{DFE receiver architecture of a D-MIMO system with $M$ links and $L$ channel taps.}\label{Block diagram of ML-DFE}
\end{figure}

D-MIMO can potentially increase  the data rate at the price of an additional ILI. Equalization plays an important role in D-MIMO systems to mitigate both ISI and ILI. D-MIMO Decision Feedback Equalization (DFE) has the benefit of mitigating the effects of the diffusive channel memory by employing one-shot detectors. Fig. \ref{Block diagram of ML-DFE} shows the general configuration of the D-MIMO receiver based on the knowledge of the CSI (Sect.III). D-MIMO DFE consists of a feedback unit that mitigates the channel memory paired with a linear filter $\mathrm{\vect{T}}$ that removes the ILI.

Let the received signal be 
\begin{equation}\label{y plus noise}
\vect{y}[k]= \vect{\bar{y}}[k]+ \vect{\omega}[k]
\end{equation}
where $\vect{\bar{y}}[k]=\vect{\bar{C}} \, \vect{X}[k]$ as in Eqn. (\ref{D-MIMO k relation}), and 
\begin{equation}\label{omega k}
\vect{\omega}[k]=(\vect{C}-\vect{\bar{C}}) \vect{X}[k] 
\end{equation}
is the stochastic part that accounts for the deviation from the mean. The conditional moments of $\vect{\omega}$ are 
\begin{equation}
\mathbb{E} \, \{\vect{\omega}[k] \}=0,
\end{equation}
\begin{equation}\label{covariance omega}
\vect{C}_{\vect{\omega} | \vect{X} } = \mathrm{cov}\{ \vect{\omega}[k] | \vect{X}[k] \} =\mathrm{diag} \{ \vect{\bar{y}}[k] \}
\end{equation}
where the last equality follows from Poisson properties after some algebra. We can further average the $\vect{C}_{\vect{\omega} | \vect{X} }=\textit{diag} \{ \vect{\bar{C}} \vect{X}[k] \}$ to have unconditional covariance:
\begin{equation}
\vect{C}_{\vect{\omega}}= \mathbb{E}_{\vect{X}} \{ \vect{C}_{\vect{\omega} | \vect{X}} (\vect{X}) \}=\mathrm{diag} \{ \vect{\bar{C}} \vect{P} \},
\end{equation}
where $\vect{P}$ is a $(ML+1)\times 1$ vector that contains $p=\mathrm{Prob}\{x_i=1\}$ as elements: $\vect{P} = p \vect{1}$. Given the model (\ref{y plus noise}) containing the D-MIMO channel mean term, and the stochastic term (\ref{omega k}), the design of the linear filter $\vect{T}$ depends on the optimization criteria for the one-shot detectors.

 \subsection{D-MIMO DFE}
 Feedback unit uses the previously decoded bits to remove the mean value of causal ISI and ILI from the observation vector $\vect{y}[k]$: $\vect{y}^*[k]= \vect{y}[k] - \vect{y}_{\mathrm{DFE}}[k]$ where
\begin {equation}\label{feedback}
\vect{y}_{\mathrm{DFE}}[k]=
\sum_{\ell=1}^{L-1} \bar{\vect{C}}[\ell] 
\vect{x}[k-\ell]  + \bar{\vect{v}},
\end{equation}
and $\vect{y}^*[k]$ is the $M\times 1$ equalized data at $k$-th time interval 
\begin{equation}\label{dfe_final}
\vect{y}^*[k]=\bar{\vect{C}}[0] \vect{x}[k]+\vect{\omega}[k].
\end{equation}
The ILI/ISI cancellation in this context  is the \textit{DFE equalizer in the mean} as it only removes the mean value of Poisson distributed interference due to the channel memory in (\ref{dfe_final}), and the stochastic part $\vect{\omega}[k]$ remains untouched after equalization because feedback interference cancellation acts on the mean values. The term $\vect{y}^*[k]$ is still suffering from  D-MIMO crosstalk with a zero-mean stochastic term $\vect{\omega}[k]$ that can be accounted for the design of linear filter $\vect{T}$. 

 We highlight that bit interval time $T_{int}$ should be  greater than diffusive pulse delay $\tau_{max}$: $T_{int} \geq \tau_{max}$ to guarantee the causality of the system for DFE. Diffusive pulse delay,  or peak time $\tau_{max}=d_{ii}^2/6D$, is the time instant corresponding to the maximum of the received number of molecules after their instantaneous release. If choosing $T_{int} < \tau_{max}$, the response is non-causal and  mitigation in DFE would be incomplete.

\subsection{D-MIMO ZF-DFE and MMSE-DFE}
The Poisson distribution of $\vect{\omega} [k]$ can be approximated by  a Gaussian one when the number of received molecules (\ref{y plus noise}) is large enough \cite{marcone2017gaussian}, so that 
$\vect{y}^*[k]\sim \mathcal{N} (\bar{\vect{C}}[0] \, \vect{x}[k],\vect{C}_{\vect{\omega}})$. Equalizers depend on the  choice of the  filter $\vect{T}$ as shown in Fig. \ref{Block diagram of ML-DFE}.  Similarly to the conventional MIMO equalizers \cite{al2000finite,spagnolini2018statistical}, ZF-DFE and MMSE-DFE can be extended to D-MIMO MC systems for the choice of filtering $\vect{T}$ summarized in Table \ref{my-label}. 
The decision
\begin{equation}
\vect{x}^*[k]=\vect{T} \,\, \vect{y}^*[k]  
\end{equation}
is for one-shot detection 
\begin{equation}\label{comparator}
\setstacktabbedgap{1pt}
 \hat{x}_i^{\mathrm{t}} [k]= 
 \begin{cases}
 1, & \text{if} \, \, \,\,\,\,\,   x_i^*[k]\geq \xi^t \cr
 0, & \,\, \mathrm{otherwise} 
 \end{cases}
 \end{equation}
where the threshold $0<\xi^t<1$ for $t\in \{ \mathrm{ZF, MMSE} \}$ can be calibrated for minimum error probability by numerical searches (Section \ref{Performance Evaluation and parameter study}). 
The computational complexity for ZF-DFE and MMSE-DFE detector is linear in block length $B$ and channel taps $L$, but quadratic in $M$: $\mathcal{O} (B \,L\, M^2)$.
\begin{table}
\centering
\caption{DFE Precoding matrix $\vect{T}$ }
\label{my-label}
\begin{tabular}{|l|l|l|}
\hline
\multicolumn{1}{|c|}{\textit{{Equalizer}}} & \textit{\textbf{ZF-DFE}} & \textit{\textbf{\,\,\,\,\,\,\,\,\,\,\,\, MMSE-DFE}} \\ \hline
\textit{{Choice of T}}                    & \textit{\textbf{$\bar{\vect{C}}^{-1}[0]$}}   & \textit{\textbf{$\vect{\bar{C}}^T[0] (\vect{\bar{C}}[0]\vect{\bar{C}}^T[0]+\vect{C}_{\vect{\omega}})^{-1}$}}     \\ \hline
\end{tabular}
\end{table}

\subsection{Least-Squares DFE Detector  }
Least-Squares (LS) detector finds the vector $\vect{\hat{x}}[k]=[\hat{x}_1[k],..., \hat{x}_M[k]]^T$, $\hat{x}_i[k] \in \{0,1 \}$, that minimizes the sum of square errors based on the equalized observation vector $\vect{y}^*[k]$:
\begin{equation}\label{LSD detection}
 \hat{\vect{x}}^{\tiny LS} [k]= \argminF_{\vect{x}_i[k] \in \{ 0,1\} } \, \big\|  \vect{y}^*[k]- \bar{\vect{C}}[0] \, \vect{x}[k] \big\| ^2.
 \end{equation}
The LS-DFE detector replaces the linear filtering $\vect{T}$ by the search among all possible combinations of $\vect{x}$ for the binary one that minimizes the error. LS-DFE computation complexity is exponential in M: $\mathcal{O} (B \, L\, 2^M)$. 

In a severe interference environment, ZF-DFE and MMSE-DFE have higher error probability, and  LS-DFE has a better performance at the price of higher computational complexity when $M$ is large (Section \ref{Performance Evaluation and parameter study}).

\section{D-MIMO Time Interleaving Encoding } \label{Interference Mitigation Technique}

D-MIMO MC systems suffer severely from the ILI in addition to the ISI. We have seen in Section IV that DFE equalizer mitigates the interference induced by the channel memory, leaving the D-MIMO cross-talk which  degrades  the MC performance. 

\begin{figure}
  \includegraphics[width=3.3in]{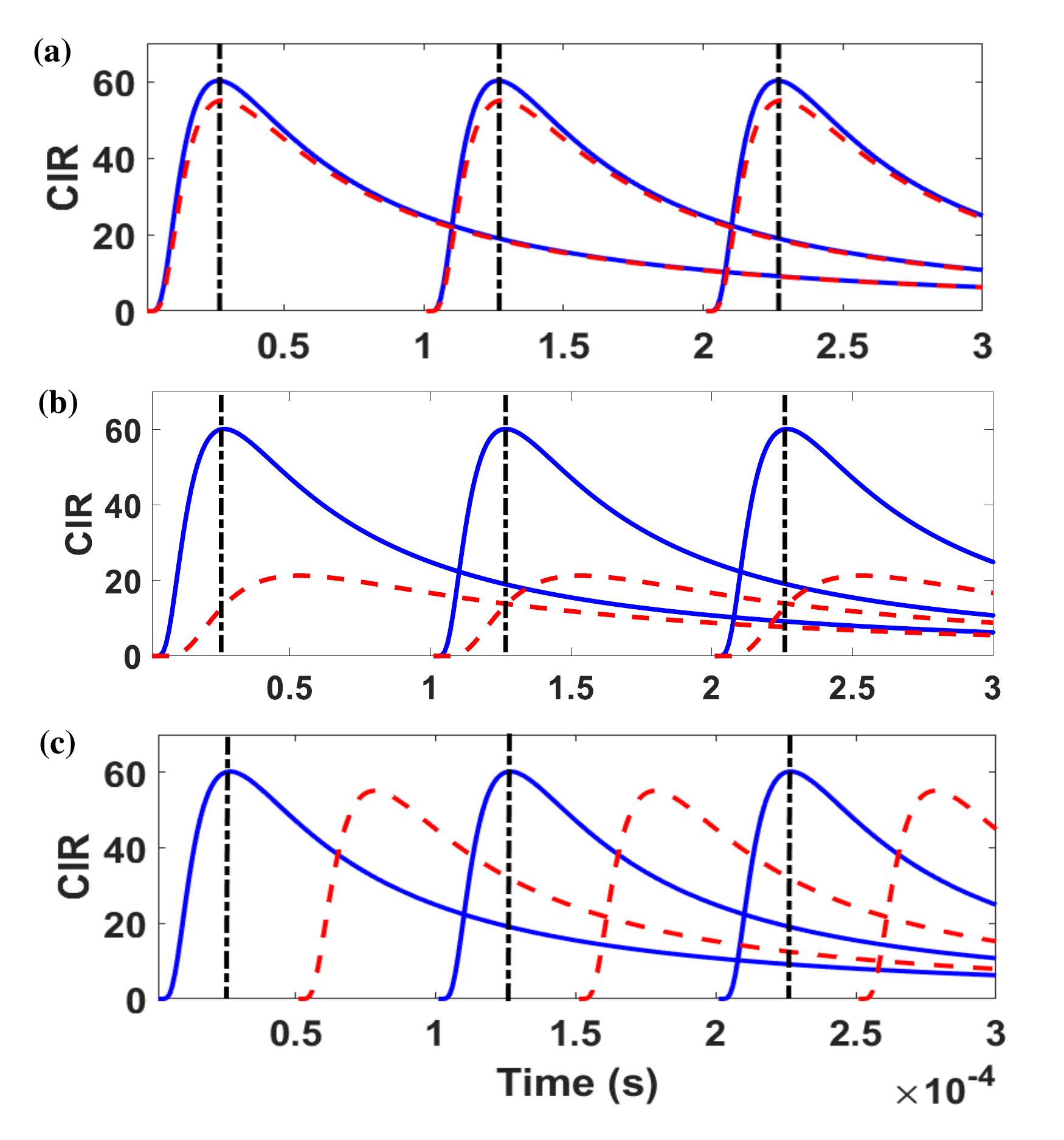}
  \centering
  \caption{CIR of a $2 \times 2$ D-MIMO system at $Rx_1$ when $d= 400\, nm$, $T_{int}=0.1 \,ms$ and (a) $h=100\,nm$, (b) $h=400\,nm$.  (c) $h=100\,nm$ and $Tx_2$ transmit with an offset time equal to $T_{off}=T_{int}/2$ respect to the $Tx_1$. Solid lines refer to the CIR of the corresponding transmitter $Tx_1$ and dashed lines refer to the CIR of $Tx_2$ which is considered as ILI.  } \label{MIMO_time diversity}
\end{figure}
D-MIMO time interleaving (TIL) encoding can easily reduce D-MIMO crosstalk by avoiding the simultaneous release of molecules at different gates of a transmitter. TIL encodes the transmission such that each gate releases the molecules at different times over each bit interval. CIR estimation for the proposed TIL encoding is the same as before, as there is no further constraint on CIR except for timing alignments at receivers. The design parameter is the interleaving offset time $T_{off}$ which can be optimized according to the D-MIMO configuration.

Fig. \ref{MIMO_time diversity} illustrates the CIR for a $2 \times 2$ D-MIMO system with $d= 400 \, \mathrm{nm}$, $T_{int}=0.2 \,\mathrm{ms}$ and $N=10^5$. Fig. \ref{MIMO_time diversity} (a) refers to the case when transmitters' inter-distance is $h=100 \,\mathrm{nm}$ where ILI is very severe. ILI  can be reduced by putting transmitters far from each other and  Fig. \ref{MIMO_time diversity} (b) refers to the case when $h=400 \, \mathrm{nm}$.  In this case, ILI component is attenuated by the loss as  the  gates' inter-distance $\mathrm{h}$ is equal to the transmitter-receiver distance $\mathrm{d}$. Fig. \ref{MIMO_time diversity} (c) refers to the case when $h=100 \,\mathrm{nm}$ and $Tx_2$ transmits with an interleaving offset time equal to $T_{off}=T_{int}/2$ respect to the $Tx_1$ and it reduces the ILIs.  It can be seen that for D-MIMO configuration where communication distance is greater than the gates inter-distance $\mathrm{d}>\mathrm{h}$, TIL encoder reduces the ILI and improves the performance (Section \ref{Performance Evaluation and parameter study}).

\begin{figure*}
  \includegraphics[width= 0.92\textwidth]{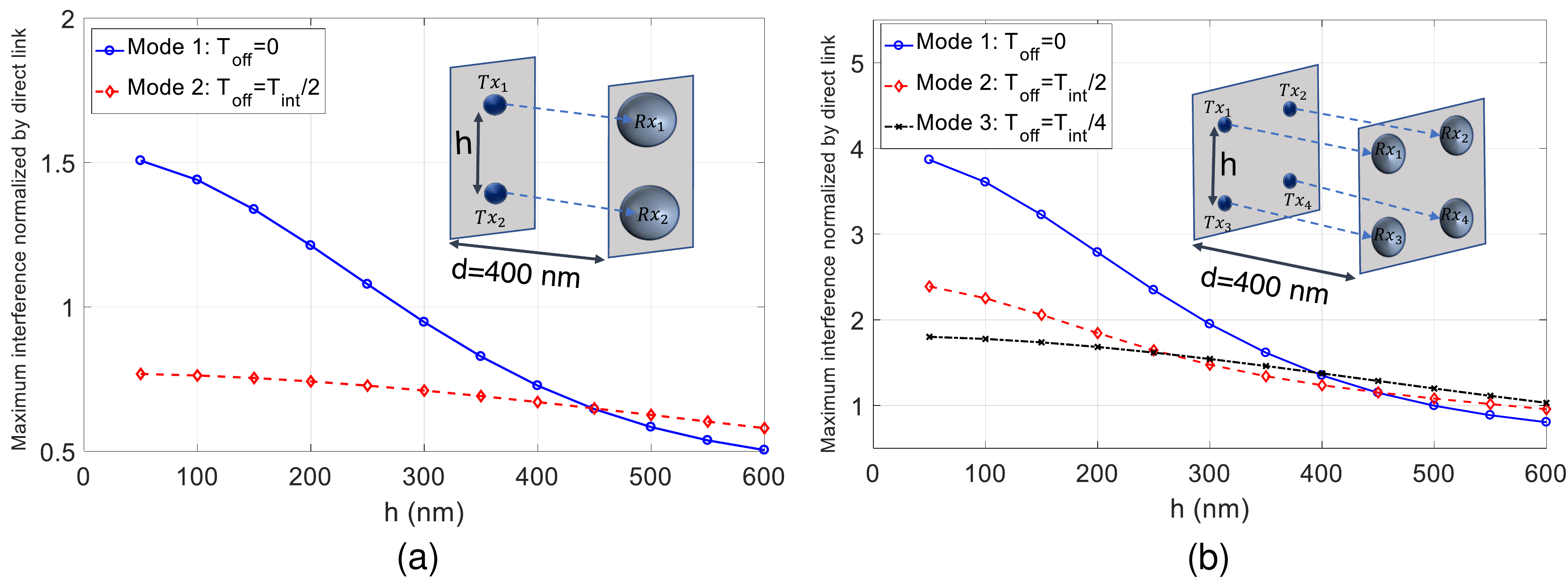}
  \centering
  \caption{ Maximum normalized mean interference
 vs.  $h$ for a $2\times2$ and $4\times4$ D-MIMO system with $T_{int}=0.2 \,\mathrm{ms}$ and $N=10^5$. } \label{MIMO_interference}
\end{figure*}
Fig. \ref{MIMO_interference}  shows the maximum normalized mean interference $(\vect{\bar{C}}_i \, \vect{1}-\bar{c}_{ii}[0])/\bar{c}_{ii}[0]$ vs. gates' inter-distance $h$ for a $2 \times 2$ and a $4 \times 4$ D-MIMO system with $d=400\,\mathrm{nm}$ and $T_{int}=0.2 \,\mathrm{ms}$. In Fig. \ref{MIMO_interference} (a) Mode $1$ refers to the case when both transmitters are releasing molecules simultaneously at the beginning of each bit interval time, while mode $2$ refers to the case when $Tx_2$ is transmitting with an offset time equal to $T_{off}=T_{int} / 2$ respect to the $Tx_1$. Notice that  receivers are assumed  synchronized with their corresponding transmitters, meaning that molecules counting time is when the CIR of the corresponding transmitter is maximum.  Fig. \ref{MIMO_interference} (b) shows the maximum normalized mean interference vs. $h$ for  a $4 \times 4$ D-MIMO system. Here, mode $2$ refers to the case when $Tx_1$ and $Tx_3$ transmit simultaneously at the beginning of bit interval time  and $Tx_2$ and $Tx_4$ transmit simultaneously  with an offset time equal to $T_{off}=T_{int} / 2$, and mode $3$ refers to the case when $Tx_1$ transmit at the beginning of the bit interval time and each transmitter is activated with an offset time equal to $T_{off}=T_{int} / 4$ with respect to each other. The remarkable interference reduction of  D-MIMO TIL encoding can be easily appreciated for both  $2 \times 2$ and $4 \times 4$ D-MIMO systems.

\section{Performance Evaluation and parameter study}\label{Performance Evaluation and parameter study}  
In this section a $2\times 2$ D-MIMO configuration is studied for OOK signaling. CIR is generated according to the analytical model  (\ref{diff concentration})-(\ref{CIR integral})  for MC systems. Diffusion coefficient for signaling molecules is  $D=10^{-9} \,\mathrm{m^2/s}$ and it is calculated according to Einstein relation \cite{nakano2013molecular} for a molecule of radius 2.4 \AA  \, in water with constant viscosity at  room temperature. Receptors are spherical shaped with radius $50\, \mathrm{nm}$ and they are mounted on the receiver nano-machines to collect the molecules. Overlap of fluctuating molecular receptors reduce the efficiency of the D-MIMO communication as
the D-MIMO channel cascades with an equivalent D-MIMO system that models the molecular capturing phenomena. The analysis of the cascade of D-MIMO is an open subject. Here we assume that receptors never overlap and this ensures that measured molecules are independent, and are transparent for the sake of complexity of the numerical analysis. The nominal distance between a paired transceivers $d_{ii}$ is known, transceivers are synchronized and the sampling time is set to the $\tau_{max}= d_{ii}^2/6D$.

Transmitters release $N$ molecules and the receivers count once the number of molecules  per each symbol. The number of channel taps for both ISI and ILI links is fixed to $L=3$. In severe interference environment (e.g. small bit interval time $T_{int}$) 
$L=3$ channel taps can not account for the long tail of diffusion, and thus it augments the noise due to truncation:
\begin{equation} \label{noise eqn}
\bar{v}_i=p\,\times \sum_{j=1}^{M}\sum_{\ell=L+1}^{L'}  \bar{ {c}}_{ij} [\ell] \,\, + \, \bar{v}_{ex},
\end{equation}
where $L'$ is an arbitrarily large number so ${c}_{ij} [L']$ is negligible. The second term $\bar{v}_{ex}$ is the noise molecules from external sources or any unintended transmitters. For simplicity, the mean of external noise is chosen as $\bar{v}_{ex}=0.05 \, \bar{c}_{ii}\, [0]$.

\subsection{Channel Estimation}
It is assumed that the nominal distance between transmitter and receiver is $d= 400 \, nm$ and the transmitter inter-distance is $h=200 \, nm$.  the static diffusive channel (i.e., no fluctuation of Tx-Rx geometry) is calculated according to  equations (\ref{diff concentration}-\ref{CIR integral}-\ref{noise eqn}):  $\vect{\bar{C}}[0]=[60.21, 41.58  ; 41.58,60.21 ] $,  $\vect{\bar{C}}[1]=[9.11 ,   8.71  ; 8.71, 9.11] $,  $\vect{\bar{C}}[2]=[3.83 ,  3.74; 3.74, 3.83  ] $ and $\vect{\bar{v}}=[  10.29; 10.29 ].$ 
    The training sequences are designed  according to (\ref{sequence}) with the smaller length $K_1=16$ and these are $\vect{s}_1=[1,1,1,0,0,0,0,1,0,1,0,1,1,0,0,1]^T$ and $\vect{s}_2=[1,1,1,0,1,0,0,0,1,1,1,0,0,0,0,1]^T$, respectively. Longer training sequences are constructed by concatenating these training sequences (as detailed in Section \ref{D-MIMO Channel Estimation}) as integer multiple of $K_1=16$.  
\begin{figure} 
  \includegraphics[width=1.05 \linewidth]{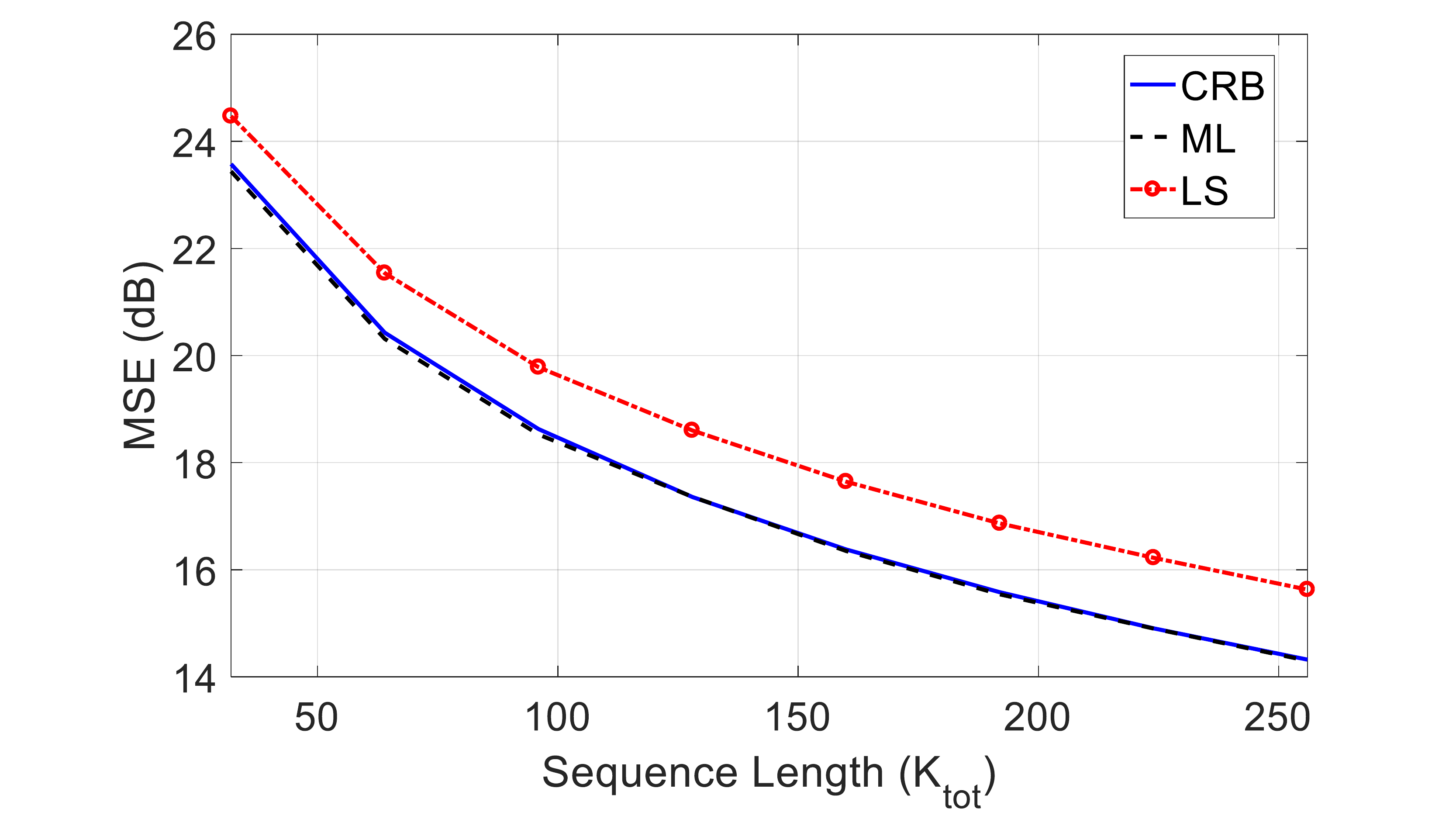}
  \centering
  \caption{MSE for CIR estimators:  ML, LS and CRB vs. the training sequence length $K_{tot}$ for a $2\times 2$ D-MIMO system with $L=3$ for a static channel.}\label{ML_LS}
\end{figure}       
In this MC configuration, each receiver has to estimate $LM+1=7$ variables, for a total of 14 variables. The results in Fig. \ref{ML_LS}, are Monte Carlo simulations with $10^4$ realization.  
Fig. \ref{ML_LS} shows the  mean square error (MSE), $\mathbbm{E} \, \{ ||  \hat{\bar{\vect{C}}} -\bar{\vect{C}} ||^{\, 2}  \} $ in dB vs. the training sequence length ($K_{tot}$) for the ML and LS CIR estimators. The MSE decreases by increasing the training sequence length as expected. Training sequences are designed to minimize the corresponding CRB. Even if ML CIR estimator uniformly outperforms the LS estimator by approximately 1 dB, the LS estimator is preferred for its simplicity to comply with the limited computational complexity of bio-nano-machines. In applications where receivers send the data to an external computing unit, the ML estimator is preferred as it attains the CRB.

\begin{figure*}
  \includegraphics[width=1.1\linewidth, left]{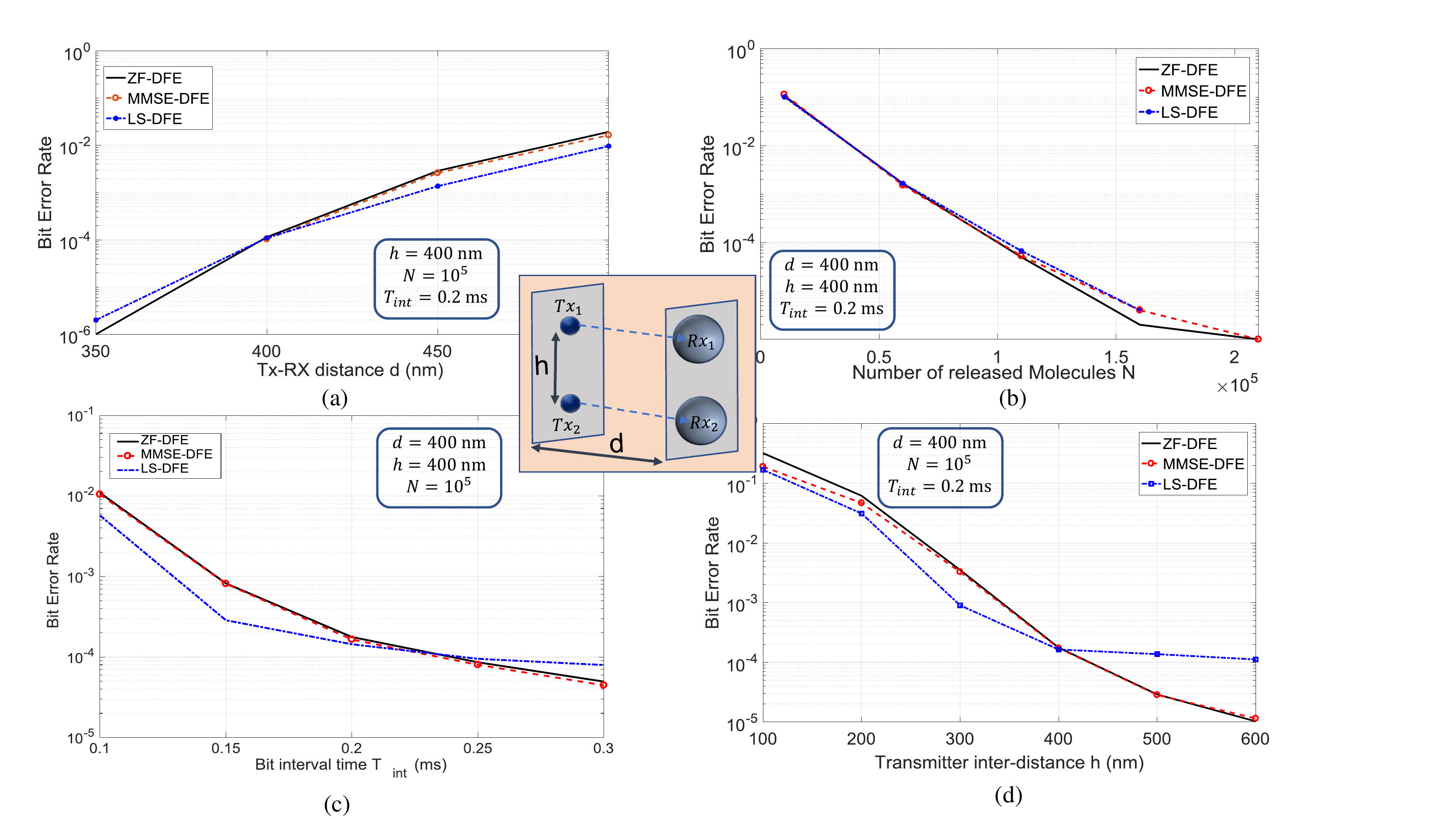}
  \caption{Performance of a $2\times 2$ D-MIMO MC system shown in the Fig. in terms of BER vs. (a) transmitter-receiver distance $d$ (b) number of released molecules $N$  (c) Bit interval time $T_{int}$ (d) transmitter inter-distance $h$.}\label{detector}
\end{figure*}
\subsection{Detection with known CSI}
Numerical results for detectors are  for a $2\times 2$ D-MIMO system. The mean CSI is known and true one,  and transceivers are fixed (static diffusive channel).  Fig. \ref{detector} shows the parametric study for different receivers presented in this paper which reveals the impact of D-MIMO system parameters on MC performance in terms of bit error rate (BER). Threshold values for ZF-DFE and MMSE-DFE equalizers are found through numerical searches and it is chosen to minimize the error probability: $\xi^{ZF}=\xi^{MMSE}=0.4$.

Specifically, 
Fig. \ref{detector} (a) compares the performance of the D-MIMO system 
in terms of error probability vs. transmitter-receiver distance $d$. The error probability  increases with $d$ as molecule loss increases.
LS-DFE outperforms ZF-DFE and MMSE-DFE in severe interference.
Notice that all three detectors are  sensitive to the number of received molecules, and when there are not enough molecules, system performance degrades drastically.

Fig. \ref{detector} (b) compares the receivers in term of error probability vs. number of transmitted molecules $N$. When more molecules are transmitted, more molecules reach the receiver and consequently performance improves and BER decreases. 

Fig. \ref{detector} (c) compares the error probability vs. bit interval time $T_{int}$. The increasing of the bit interval  time $T_{int}$ does not affect the $\vect{\bar{c}}_{ii}[0]$ (direct link), but it decreases the interference and consequently it improves the performance. When $T_{int}> 0.3 \, \mathrm{ms}$ the performance is dominated by the noise and the stochastic nature of diffusion.  Notice that here for fair comparison ISI/ILI length for simulating the received number of molecules is adapted to variations of $T_{int}$ while keeping the number of channel taps L=3  fixed for equalization and detection.

Fig. \ref{detector} (d) shows the performance of the D-MIMO system vs.  gates' inter-distance $h$. For small $h$ values, the interference from non-corresponding gates are too severe and the BER is high. Increasing $h$ makes the ILI decrease and the performance improves. 
One can reduce the distance between gates of each transmitter to reduce the total size of the nanomachines, and thus maximize the data rate per unit of space. Therefore, in the next subsection D-MIMO TIL encoding is employed to mitigate the D-MIMO crosstalk for small gates' inter-distance $h$. 
 
It can be seen in Fig. \ref{detector} that in low interference environment ZF-DFE and MMSE-DFE outperforms the LS-DFE as they employ equalization which decrease error probability. However, in severe interference environment, the equalization is not effective and LS leads to lower error probability at the price of exponential computational complexity in $M$.

\subsection{D-MIMO Time Interleaving Encoding}
\begin{figure} 
\includegraphics[width=1\linewidth]{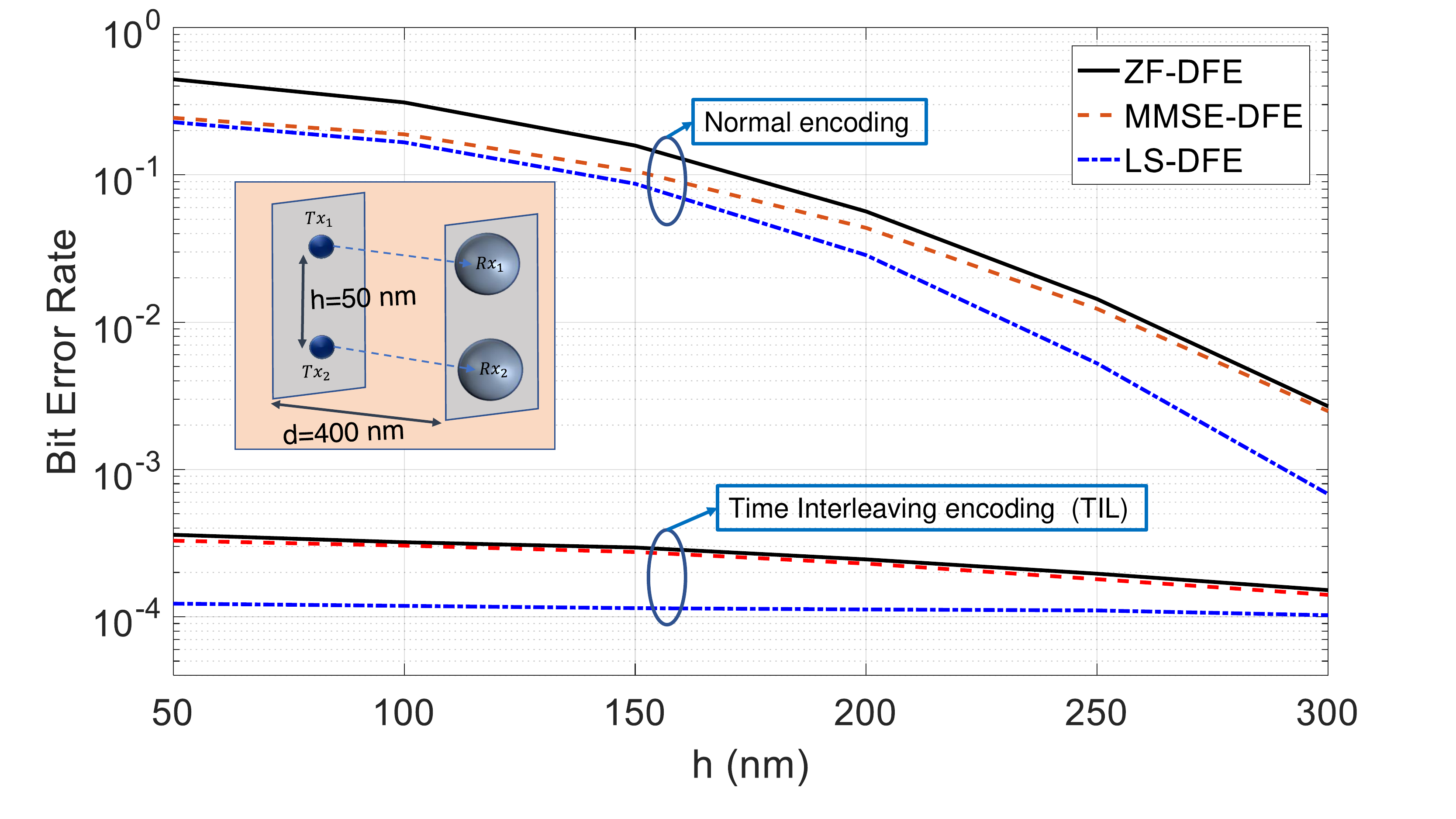}
\centering
\caption{Performance of the $2 \times 2$ D-MIMO system with $N=10^5$ and $T_{int}=0.2 \, \mathrm{ms}$ shown in the figure in terms of  BER vs. transmitters inter-distance $h$ for normal encoding and time interleaving (TIL) encoding at the transmitter.}\label{performance TIL} 
\end{figure}

D-MIMO TIL encoding reduces the ILI when the Tx-Rx distance is larger than gates' inter-distance $d>h$. Fig. \ref{performance TIL} shows the numerical results for the $2\times 2$ D-MIMO system for both normal and TIL encoding at the transmitters. It can be seen that
for small $h$, say $h\le  100 \, \mathrm{nm}$, interference is severe and BER might be high. On the other hand, D-MIMO TIL encoding mitigates the MIMO cross-talk and improves the BER such that it is approximately independent of $h$ as the ILI is mitigated. BER floor is due to the residual ISI affected by the stochastic nature of diffusion. The LS-DFE outperforms the ZF-DFE and MMSE-DFE receivers, because it does not rely on Gaussian approximation. 

 \subsection{Block-Type Communication (static)}
  
The block-type D-MIMO processing for MC system is evaluated here for a static diffusive channel. Transmitters send the training sequences designed in Section \ref{D-MIMO Channel Estimation} through the diffusive channel at the beginning of each block, the receiver estimates the CIR by knowing the training sequence and counting the received molecules on the information bits. DFE cancels the mean interference based on the estimated CIR and one-shot detectors decode the information (Fig. \ref{Block diagram of ML-DFE}). The accuracy of the estimated CIR is related to the length of training sequence (Fig. \ref{ML_LS}). The performance of the system would reach to the ideal case where true  CSI is known if accuracy of the estimated CIR is good enough.

\begin{figure} 
\includegraphics[width=1 \linewidth]{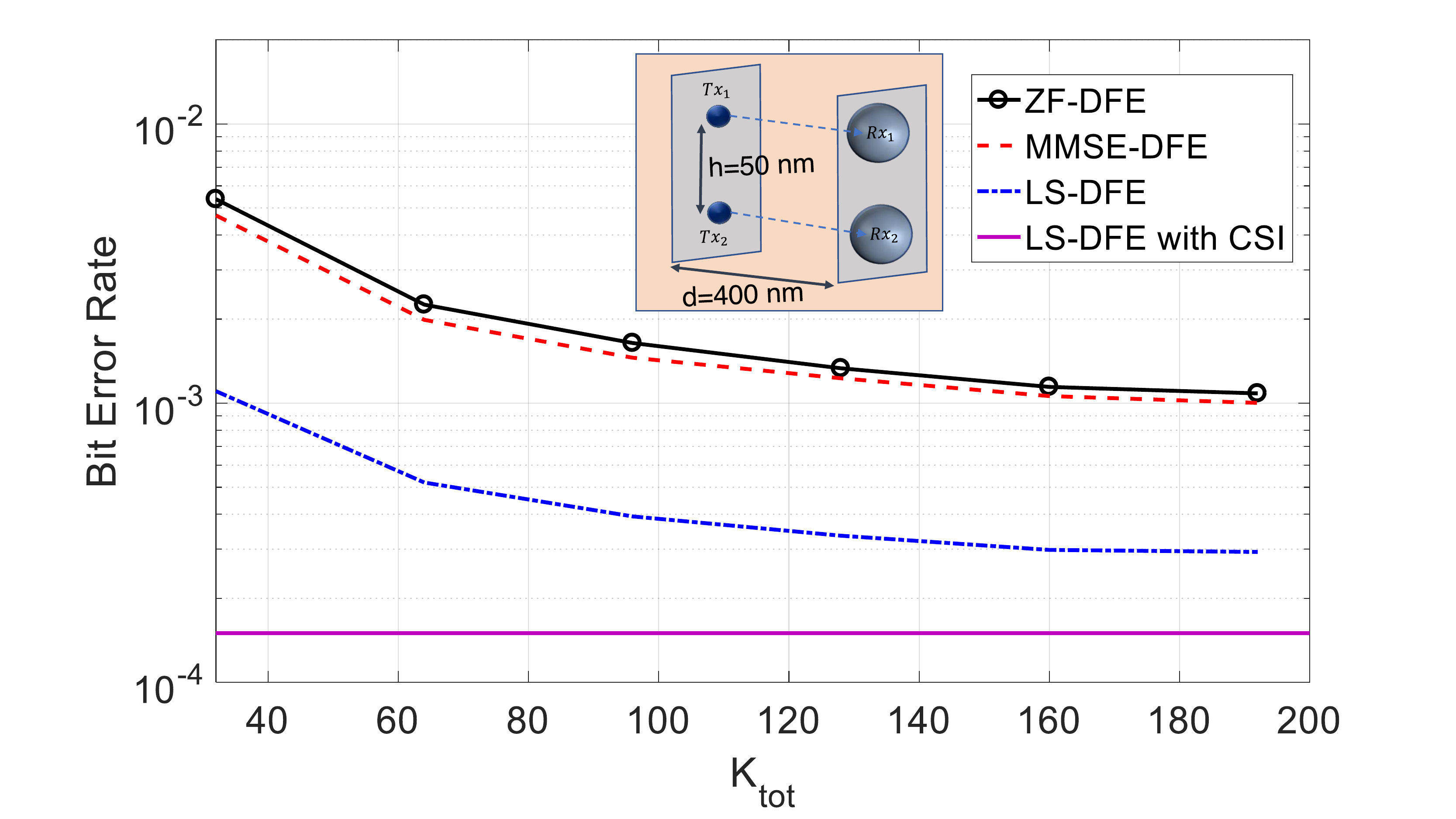}
\centering
\caption{Error probability of the $2 \times 2$ D-MIMO system with $N=10^5$ and $T_{int}=0.2 \, \mathrm{ms}$ vs. total training sequence length for LS CIR estimator. Performance of LS-DFE with perfectly known CSI is shown for the sake of comparison.  } \label{performance Ktot} 
\end{figure}
The total number of bits in each block is $B$, a total number of pilot bits $K_{tot}=M\times K$ are used for CIR estimation and the remaining $B-K_{tot}$ are used as information bits. Fig. \ref{performance Ktot} shows the error probability vs. training sequence length $K_{tot}$ for the same setting in Fig. \ref{performance TIL} . It can be seen that as $K_{tot}$ increases, the BER decreases and for $K_{tot}>128$ the BER decreases smoothly.
The price for CIR estimation is $K_{tot}$ out of the B block information  and small loss of performance.

\subsection{Block-Type Communication (time-varying) }
A time-variying diffusive channel is considered herein where transmitter and receiver are moving randomly and their mobility is modelled by Brownian motion \cite{ahmadzadeh2017diffusive,ahmadzadeh2018stochastic}. We have assumed that transceivers are hard spheres with radius $r_x$ and their diffusion coefficient $D_X$ for movement is calculated according to Stokes-Einstein relation \cite{nakano2013molecular}. It is assumed that channel is quasi-static during channel coherence time $T_c=n\times T_{int}$ and n is a positive integer. The case $T_c= 1\times T_{int} $ means channel varies after every symbol transmission. We remark that transceivers are also moving during channel estimation phase and therefore CIR estimation is averaged over channel variations.  
Particularly, this section investigates the effect of the channel variations on the  performance of a block-type D-MIMO MC communication.

Three-dimensional random walk is assumed as follows: transceivers are at the initial positions  at the beginning of each block: $P_{Tx_1}(t=0)=(0,0,0)$, $P_{Tx_2}(t=0)=(0,h,0)$, $P_{Rx_1}(t=0)=(d,0,0)$, $P_{Rx_2}(t=0)=(d,h,0)$,  and then they move randomly during the block: $P_{Tx_j}(t=q\, T_c)=P_{Tx_j}(t=(q-1)\, T_c)+ \delta_t $ and $P_{Rx_i}(t=q\, T_c)=P_{Rx_i}(t=(q-1)\, T_c)+ \delta_r $ where $\mathrm{q}$ is a positive integer and $\delta_{ t,r} = (\delta_x,\delta_y,\delta_z)$  and $\delta_{x,y,z} \sim \mathcal{N}(0, 2\, D_x \, T_c)$. Small $T_c$ implies larger number of steps with smaller step size, and large  $T_c$ implies fewer number of steps with larger step size.

Results shown in this section are by Monte Carlo simulations with $10^4$ random block transmission. Least-squares CIR estimator and ZF-DFE detector are used at receiver as they have lower computational complexity. The following parameters are assumed in this section unless it is mentioned: bit interval time $T_{int}=0.2 \, \mathrm{ms}$,  $N=5\times 10^5$ molecules are released in each transmission to signal bit 1. 

\begin{figure} 
\includegraphics[width=1\linewidth]{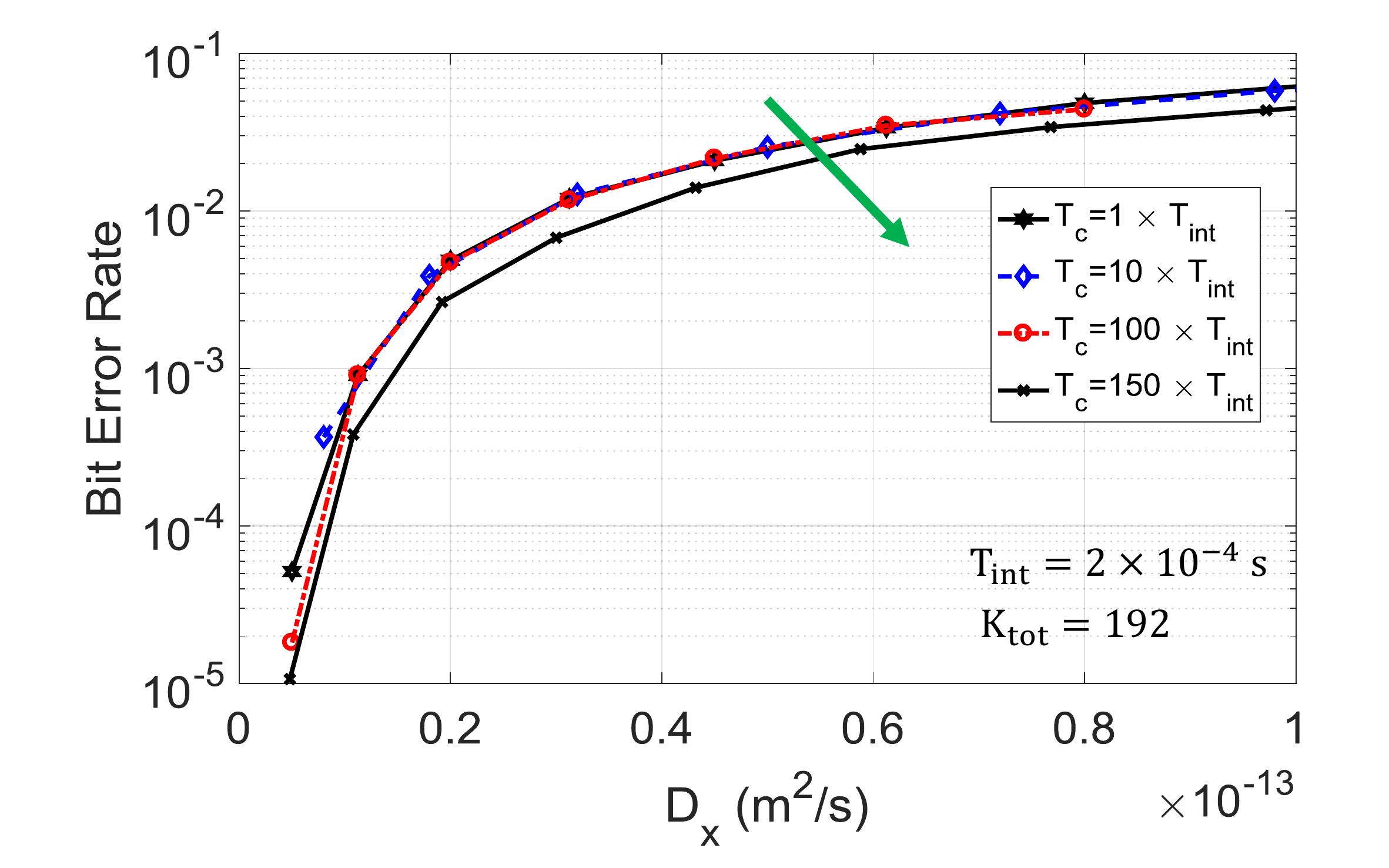}
\centering
\caption{Performance of a $2\times 2$ D-MIMO MC system with $h=50 \, \mathrm{nm}$ and ${N}=5\times 10^5$ and ${B}=600$. in terms of BER vs transceivers' diffusion coefficient ${D_X}$ for different channel coherence time  $T_c$. }\label{BERvsmu} 
\end{figure}
\begin{figure*} 
\includegraphics[width=0.95\linewidth]{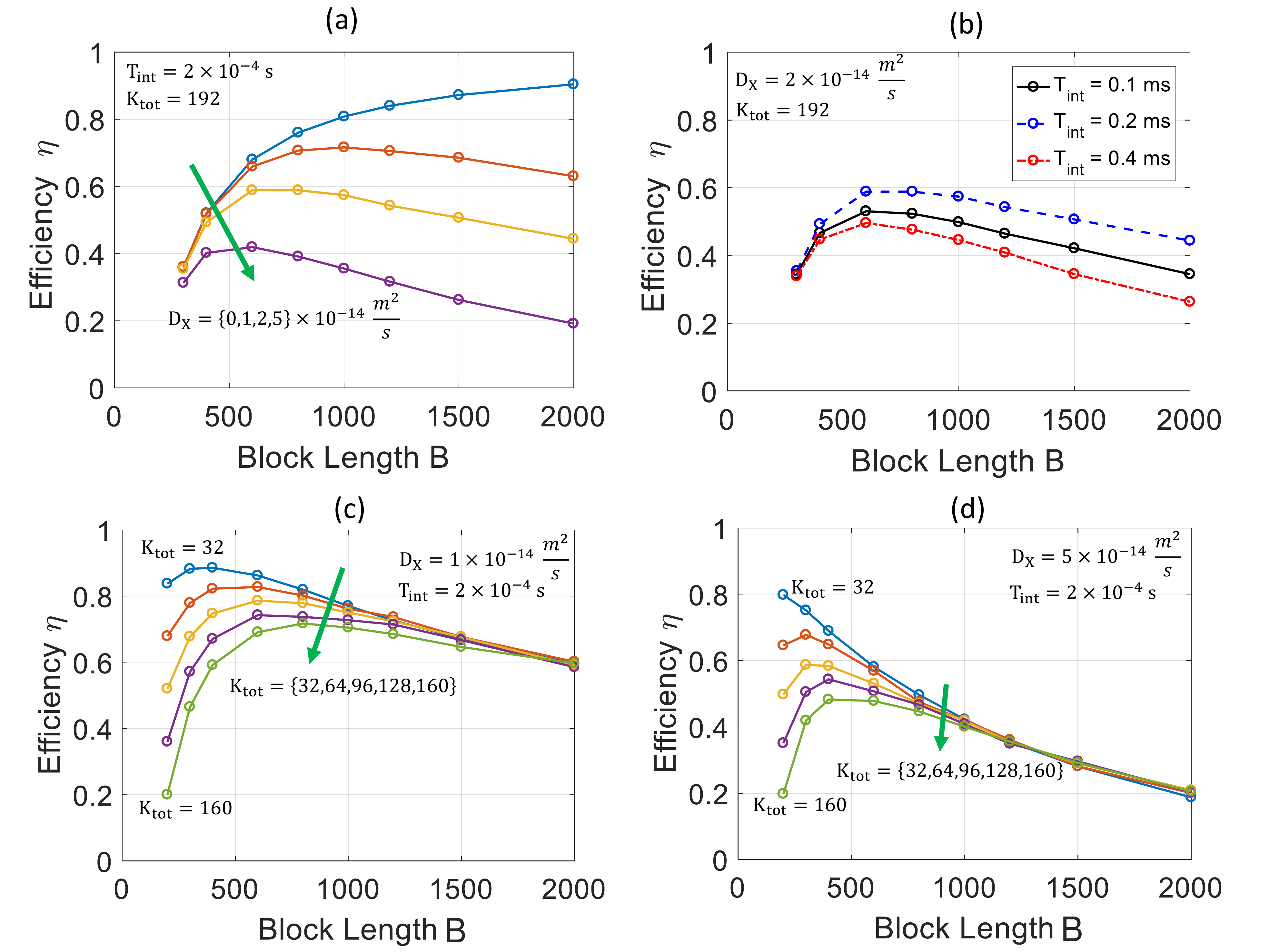}
\centering
\caption{System performance in terms of efficiency (Eqn. \ref{efficiency}) vs block length B, a) for different transceiver diffusion coefficient $D_X$,  b) for different bit interval time $\mathrm{T_{int}}$, c \& d) for different training sequence length $K_{tot}$}\label{Eta_vs_B} 
\end{figure*}

Fig. \ref{BERvsmu} (a) shows the error probability vs transceivers diffusion coefficient $D_X$. Here, error probability is the average BER over the block for block length $B=600$. It can be seen that BER depends highly on the transceivers diffusion coefficient ${D_X}$ but not on the choice of channel coherence time ${T_c}$. 
In other words, simulations are accurate as far as $T_c<<T_B=B\times T_{int}$.  For the rest of the paper, channel coherence time is considered $T_c=10\times T_{int}$ (as it leads to lower simulation complexity). 

Fig. \ref{Eta_vs_B} shows the overall performance of a system in terms of efficiency vs. block length. Here efficiency is defined as
\begin{equation}\label{efficiency}
    \eta =\frac{(B-K_{tot})}{B} \,\times (1-P_s)
\end{equation}
where $P_s$ is the packet error probability, and packet is discarded with one error within $B-K_{tot}$ bits. The efficiency $\eta$ includes either the block-efficiency due to the number of training used to estimate the channel in every block, and the error probability in the form of ratio of the number of successful decoded block of $B$-bits without any error in $B-K_{tot}$ bits out of the total blocks.

Fig. \ref{Eta_vs_B} (a) shows the system efficiency vs block length for static diffusion channel (${D_X}=0$) and transceivers diffusion coefficient ${D_X}=\{1, 2, 5\}\times 10^{-14}$ corresponding to transceivers' radius $r_x=\{24, 12, 4.8  \}\mu m$. The D-MIMO system efficiency decreases when transceivers diffusion coefficient $D_X$ is larger and one can optimize the block length for each channel condition. Fig. \ref{Eta_vs_B} (b) shows the effect of bit interval time $T_{int}$ on efficiency $\eta$. It can be seen that efficiency is higher for the choice $T_{int}= 0.2 \, \mathrm{ms}$. The choice $T_{int}= 0.4\,  \mathrm{ms}$ leads to larger error probability because the estimated CIR outdates during the block (transceivers deviate more for the same block length), and the choice $T_{int}=0.1 \,  \mathrm{ms}$ also leads to higher error probability because the interference is large.
 Fig. \ref{Eta_vs_B} (c-d) shows that efficiency increases when $K_{tot}$ is smaller. The CIR outdates faster when transceivers deviate faster (larger $D_X$) and therefore smaller training sequence and smaller block length leads to higher efficiency.

\section{Conclusions}
In this paper a pragmatic approach to design a D-MIMO MC system is proposed. Block-type communication is assumed in a time-varying diffusive channel where the CIR is estimated at the beginning of each block. ML and LS training-based CIR estimators are derived  and their performance are compared with Cramér-Rao bound. ML CIR estimator attains the CRB and outperforms the LS CIR estimator by 1 dB at the expense of higher computational complexity. For decoding, the DFE is used to mitigate the  severe ISI due to the D-MIMO channel memory. Several one-shot detectors based on DFE are discussed. ZF-DFE and MMSE-DFE outperforms LS-DFE in low interference environment due to the further equalization. However, in severe interference situation ZF and MMSE equalizers are not that effective, and LS-DFE has to be preferred at the price of higher computational complexity.  D-MIMO systems suffer severely from the ILI due to the  cross-link  coupling. D-MIMO time interleaving encoding is proposed to mitigate the the ILI and improve the error probability while preserving the bit-rate. Finally, the performance of the D-MIMO MC system assuming block-type communication is investigated in terms of throughput vs. bit interval time $T_{int}$. It is shown that for large $T_{int}$, bit error rate is small and the only loss is due to the small fraction of information bits which are used as training sequence for CIR estimation. 

Here, we have assumed transceivers move randomly and their mobility is modelled by Brownian motion and the validity of the CIR estimation in block-type communication is discussed. It has been shown that training sequence length and block length can be optimized to maximize the overall D-MIMO MC  efficiency.


\begin{IEEEbiography}[{\includegraphics[width=1.5in,height=1.2in,clip,keepaspectratio]{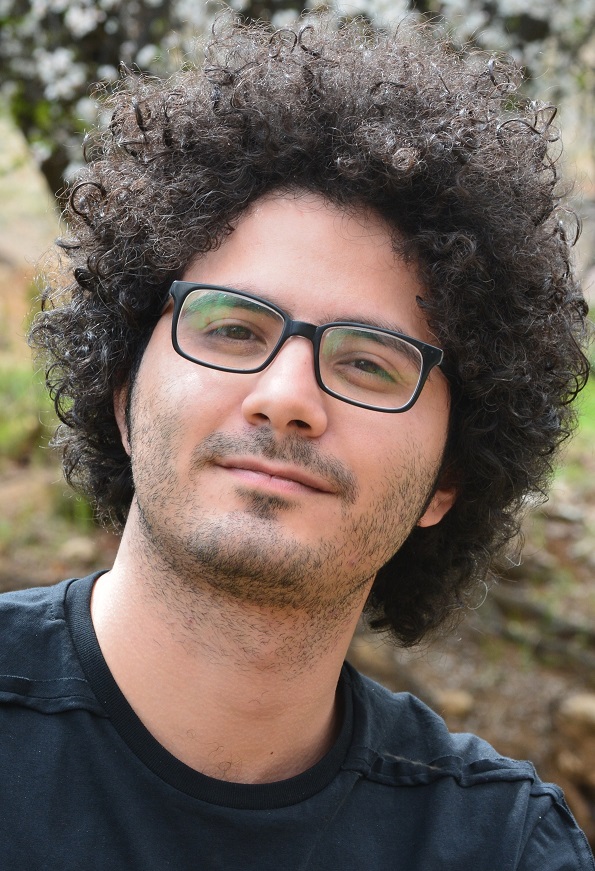}}]{Reza Rouzegar}
received the M.Sc. degree (Hons.)
in telecommunication engineering from the Politec
nico di Milano, Milan, Italy, in 2017. He is currently 
pursuing the Ph.D. degree with the THz Physics 
Group, Free University of Berlin, Berlin, Germany. 

His research interests include signal processing 
and physical layer for different communication 
schemes including optical, wireless, and molecular 
networks. He is also a member of the International 
Max-Planck Research School (IMPRS) focusing on 
surface science and ultrafast spin dynamics.
\end{IEEEbiography}

\begin{IEEEbiography}[{\includegraphics[width=1.5in,height=1.2in,clip,keepaspectratio]{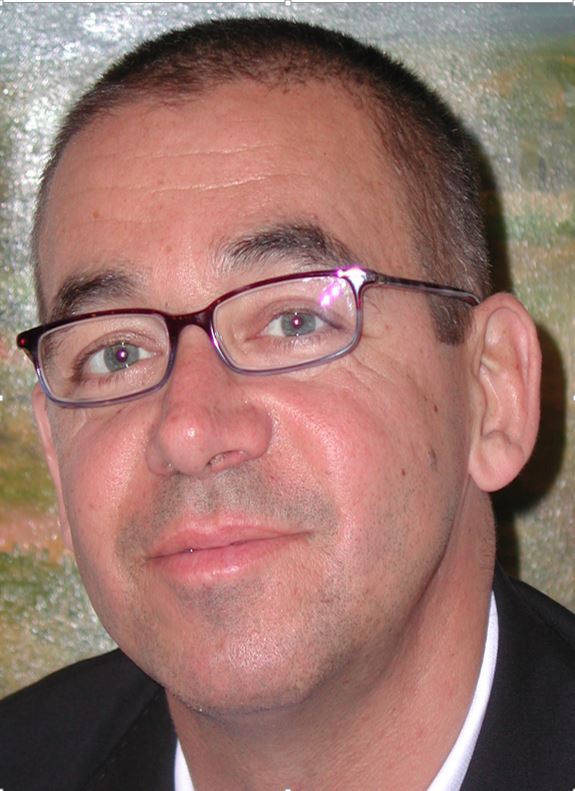}}]{Umberto Spagnolini}
(SM’03) graduated as Dott.Ing.Elettronica (cum laude) from the Politecnico di Milano in 1988. Since 1990 he has been Faculty member of the Dipartimento di Elettronica e Informazione, Politecnico di Milano, where he is Professor in Signal Processing and Telecommunications.

His research focuses on statistical signal processing, communication systems, and advanced topics in signal processing for remote sensing and wireless communication systems. Within these areas, he is author of more than 300 papers on peer-reviewed journals/conferences and patents. The specific areas of interest include channel estimation and space–time processing for wireless communication systems (GSM, UMTS, WiMAX, LTE, and now 5G), diffusive communication, cooperative and distributed systems, parameter estimation/tracking and wavefield interpolation for UWB radar, oil exploration and remote sensing. He wrote a book Statistical Signal Processing in Engineering (Wiley Ed., 2018). 

\end{IEEEbiography}

\end{document}